\documentclass[journal,comsoc]{IEEEtran}

\usepackage[T1]{fontenc}
\usepackage{cite}
\usepackage{amsmath}
\usepackage[cmintegrals]{newtxmath}
\usepackage{amsfonts,amssymb}
\usepackage{latexsym}
\usepackage{graphicx}
\usepackage{float} 
\usepackage{subfigure}
\usepackage{booktabs}
\usepackage{url}
\usepackage{multirow}
\begin{document}

%
\title{Deep Prototypical Networks Based  \\ Domain Adaptation for Fault Diagnosis}
%
%
%

\author{Huanjie~Wang,~
        Jie~Tan,~
        Xiwei~Bai~
        and~Jiechao~Yang
\thanks{Huanjie Wang, Xiwei Bai and Jiechao Yang are with University of Chinese Academy of Sciences and the Institute of Automation, Chinese Academy of Sciences, Beijing 100190, China (e-mail: wanghuanjie2018@ia.ac.cn; baixiwei2015@ia.ac.cn; yangjiechao2018@ia.ac.cn).}
\thanks{Jie Tan is with the Institute of Automation, Chinese Academy of Sciences, Beijing 100190, China (e-mail: tan.jie@tom.com).}}

\maketitle

\begin{abstract}
Due to the existence of dataset shifts, the distributions of data acquired from different working conditions  show significant  differences in real-world  industrial applications, which leads to performance degradation of traditional machine learning methods. 
This work provides a framework that combines supervised domain adaptation with prototype learning for fault diagnosis. 
The main idea of domain adaptation is  to apply the Siamese architecture to learn a latent space where the mapped features are inter-class separable and intra-class similar  for both source and target domains.
Moreover, the prototypical layer utilizes the features from Siamese architecture to learn prototype representations of each class.  
This supervised method is attractive because it needs very few  labeled target samples. 
Moreover, it can be further extended to address the problem when the classes from the source and target domains are not completely overlapping. The model must generalize to unseen classes in the source dataset, given only a few examples of each new target class. Experimental results, on the Case Western Reserve University bearing dataset, show the effectiveness of the proposed framework. With increasing target samples in training, the model quickly converges with  high classification accuracy.
\end{abstract}


%
\IEEEpeerreviewmaketitle

\section{Introduction}
%
%
%
%
\IEEEPARstart{T}he advances in  the manufacturing industry make the volume of data proliferate. Making full use of the manufacturing big data, it can  further promote industrial development. 
The sensor data that contain machine health information play an essential  role in fault diagnosis and prognosis. 
In the past decades, many fault diagnosis methods have been proposed\cite{li2000stochastic,dong2007hidden,simani2003model,lee2016convolutional,frank1997fuzzy,liu2016robust}.
Because the physics-based methods usually require specific a priori knowledge,  they are  laborious to implement as the industrial environment changes. Compared with traditional physics-based models, data-driven methods have attracted much attention due to their effectiveness and flexibility. Studies that combine  signal processing and machine learning techniques have obtained impressive results in many industrial cases\cite{aldrich2013unsupervised,lv2016fault,jegadeeshwaran2015fault,elforjani2017prognosis,martin2018experimental}. 
In \cite{verstraete2017deep}, an automatic deep feature  learning method uses time-frequency images to train a deep convolutional neural network for bearing fault{} diagnosis.
Glowacz et al.\cite{glowacz2018early} proposed a new feature extraction method called MSAF-20-MULTIEXPANDED.
The features extracted from acoustic signals of the single-phase induction motor are classified by machine learning methods.
In \cite{wang2019novel}, 
the maximum kurtosis spectral entropy deconvolution (MKSED) method  that uses the signal denoised by  Ensemble Empirical Mode Decomposition (EEMD)  is applied to classify bearing fault.

However, most of the intelligent fault diagnosis methods require a large amount of labeled data (target data) for training, which restricts their extensive applications. 
Several situations that could not obtain sufficient samples to train the deep model\cite{zhang2019limited}. 
(1) In many applications, the process from degradation to the failure  might take a long time, e.g., generators or jet engines\cite{gebraeel2009residual}. Therefore, it is difficult to collect related data.
(2) The fault detection system does not allow critical machines to operate in fault states. Once the system detects a fault, it immediately shut down the machine, which results in collecting only a few fault samples. 
(3) The working conditions frequently change in actual tasks, and the fault signal could be collected from different working conditions, even from different machines.
It is difficult to collect sufficient samples for every type of fault.
Meanwhile,  the samples of similar faults usually  show significant distribution discrepancies. It indicates that the model trained in one situation is not suitable for another. It is difficult or even impossible to recollect the new labeled  data to train a  model for the actual task.
When it is difficult to collect the target data,  the typical  approach is to use available datasets (source data) for training the related target model.
Therefore, it is vital to adapt the useful information from the source training task to a new but related diagnosis task with a few labeled target samples.

The domain adaptation, which can transfer the knowledge from the source domain to a different but related target domain, can be adopted in the situation where the source  and target data have different distributions. Although domain adaptation techniques have recently been widely used in situations such as image classification, face recognition, object detection, and so forth\cite{wang2018deep}, they have not yet been investigated thoroughly for time series classification, especially in fault diagnosis. In recent years, several fault diagnosis methods based on transfer learning have been proposed \cite{lu2016deep,guo2018deep,han2019deep}. In \cite{han2019deep}, a  framework with joint distribution adaptation (JDA) adapts the unlabeled target data 
to  the conditional distribution. In \cite{zhang2019limited}, a few-shot learning neural network based on the Siamese network is used for  bearing fault diagnosis with limited data.

In the field of fault diagnosis, the data-driven methods  mainly focus on how to use fewer  source data to learn more information or the transferability between different working conditions of the same machine. However, the amount of data is not a problem in the real-world industry, and we can  slowly accumulate more labeled data. Meanwhile, different machines could have similar types of fault, and it is expensive and  time-consuming to collect labeled data for all machines.  Therefore, we pay more attention to the adaptation ability between source and target datasets, even if these datasets were collected  from different machines. 

In this work, we introduce a supervised approach for bearing fault diagnosis based on domain adaptation. The approach requires very few labeled target samples per class in training. Even  one sample can significantly increase the model performance. Furthermore, the model trained on the source dataset can generalize to new classes that  can only be seen in the target dataset, given only a few labeled samples of each new class. To improve the robustness, we adopt the prototype learning making the unseen classes  easily distinguished. The modified network is based on the assumption that there exist prototypes that can represent  corresponding  classes in the latent space. To do this, we learn a non-linear mapping to minimize the discrepancy between source and target distributions in a latent space by neural network and take a prototype to represent the  center of each class. The classification task is then simplified to find the nearest class prototype.
The framework was verified on the standard Case Western Reserve University Bearing Datasets\cite{CWRU}, which showed that our approach is effective in fault diagnosis with very few labeled target samples.

The rest of the paper is organized as follows. Section II reviews related works about domain adaptation and prototype learning. Section III describes the problem formulation and  proposed framework. A series of experiments are carried out in Section IV. 
Finally, conclusions and future works are presented in Section V.

\section{Relation works}
\noindent \textbf{Domain adaptation.} Due to the existence of dataset bias\cite{ponce2006dataset,torralba2011unbiased,tommasi2017deeper} and data shifts (e.g., prior shift, covariate shift\cite{shimodaira2000improving}, concept shift\cite{vorburger2006entropy})   between different data sources, the model trained on one dataset is not suitable for another. Domain adaptation and transfer learning are two sub-fields of machine learning that are used to solve these problems. 

Compared with the general use of transfer learning, the  domain adaptation focuses  on how to deal with different probability distributions of  datasets. Therefore, prior methods of domain adaptation mainly try to  minimize the discrepancy between the source and target samples directly. In\cite{tzeng2014deep}, the deep transfer network uses a model with shared weights to find a domain invariant space for source and target distributions.  
Moreover, an adaptation layer measures their differences with the Maximum Mean Discrepancy (MMD)\cite{gretton2007kernel} metric.
In\cite{rozantsev2018beyond}, the two-stream architecture model considers that the weights in corresponding layers are related but not shared. Therefore, they added a weight regularizer to account for the distance between the source and target distributions. In\cite{Tzeng_2017_CVPR}, a model combines adversarial learning with
discriminative feature learning, mapping target distribution to the source feature space.
According to \cite{motiian2017unified,motiian2017few}, these methods can be divided into three categories. Among these categories, the one that finds a shared domain subspace is concerned with accounting for the assumption that source and target conditioned label distributions are similar, and there exists a classifier that can work well on both source and target distributions\cite{ben2010impossibility}. In this category, Siamese networks\cite{chopra2005learning} are suitable for minimizing the discrepancy of different domains in latent space. 
The literature\cite{motiian2017unified} uses a Siamese network to make the same class from different domain datasets as close as possible. 
Since the unseen target data is severely limited for the problem when the classes from the source and target domains are not completely overlapping, the architecture of the Siamese network is challenging  to separate the new classes from each other.
We modify the model by prototype learning, which improves model performance with a few target training samples.

\noindent \textbf{Prototype learning.} 
Through searching prototypes to represent the centers of the  data distribution in each class, prototype learning is effective in 
improving the performance of classification. The simplest method of prototype learning is  the unsupervised clustering, which searches the class centers  used as the reduced prototypes independently\cite{bezdek1998multiple,liu2001evaluation}. Since the unsupervised clustering does not consider the class information, the  classification accuracy is usually lower compared with supervised classification methods. The learning vector quantization(LVQ)\cite{kohonen1990self}, proposed by Kohonen, supervised adjusts the weight vectors based on searching the optimal position of the prototypes. Although the convergence is not guaranteed, the attractive performance makes LVQ popular in many works.  In the variations of LVQ, the parameter optimization approaches, which learn prototypes through optimizing the objective functions by gradient search, have excellent  convergence property in learning\cite{sato1996generalized,sato1998formulation}. In \cite{snell2017prototypical}, the prototypical networks  search prototype representations of each class in a metric space.
In \cite{yang2018robust}, the model, which combines the prototype-based classifiers with deep convolutional neural networks, improves the model robustness. In our study, we minimize the discrepancy of  source and target distributions by domain adaptation and learn the best representations of different classes, making the prototypes as far as possible from each other to improve the accuracy of classification. 
The main contributions of this work are summarized in the following.
\begin{enumerate} 
\item The framework uses convolutional neural networks as a basic model applied to time series classification. It can learn a domain invariant space with effective domain adaptation capacity through the Siamese architecture.
\item 
The framework can be extended to solve the problem when the classes from the source and target domains are not completely overlapping.
The model learned on the source dataset can generalize to new classes that can only be found in the target dataset, given only a few samples of each new class in training.
\item Compared with the traditional classification methods, our model adds  weight regularizations to adjust the distance of different prototypes, making the new class centers easily distinguished.
\item With  attractive robustness, disrupting the corresponding classes between source and target domains does not affect the classification accuracy, which is suitable for the complex working conditions in industrial applications.
\end{enumerate}

\begin{figure}
\centerline{\includegraphics[width=3.5in]{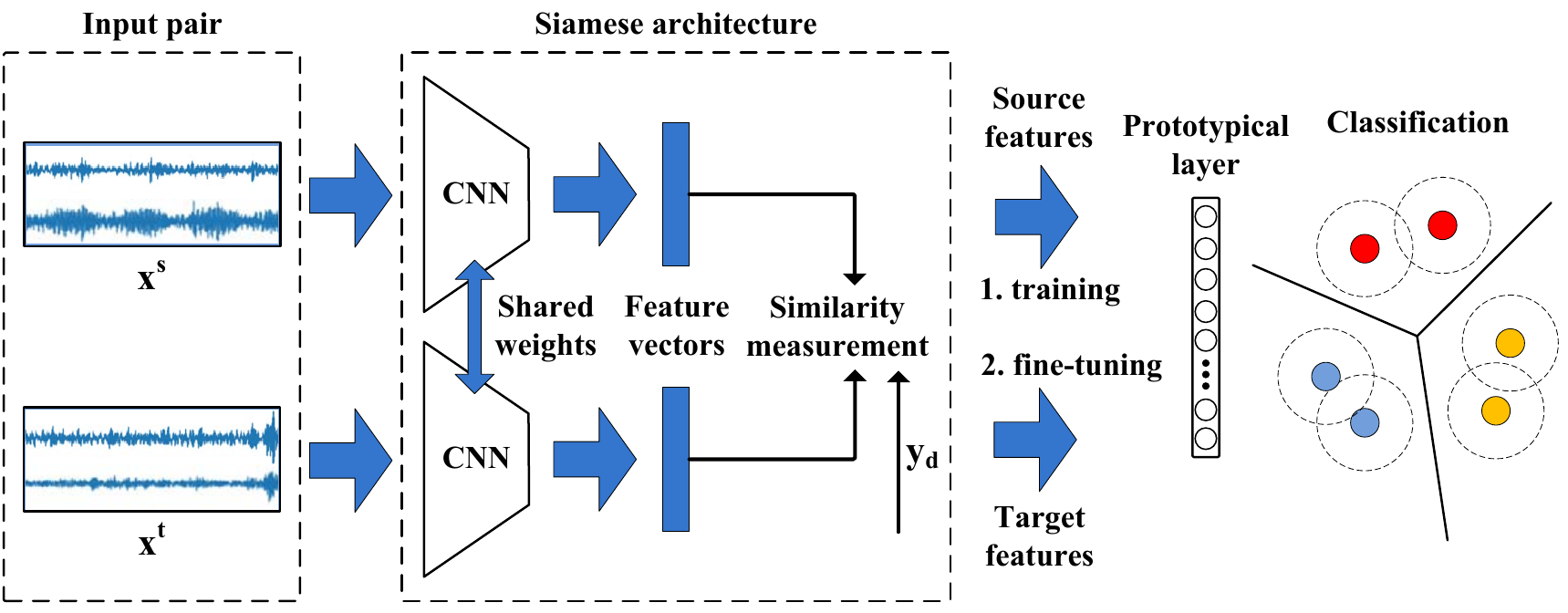}}
\caption{Deep prototypical adaptation learning:
The two input streams are two-dimensional time-series samples from source and target domains respectively. In training, the distance loss $\mathcal{L}_D$ minimizes the discrepancy between samples in the same classes from different domains. Meanwhile, the classification loss guarantees the  different classes centers easily distinguished and high accuracy. In this figure, we learn two prototypes for each class and use prototype matching for classification.
}
\label{fig1}
\end{figure}

\section{methods}

\subsection{Notation}

\begin{table}[!t]
\renewcommand{\arraystretch}{1.3}
\caption{Notations and Descriptions}
\centering
\setlength{\tabcolsep}{3pt}
\begin{tabular}{cc|cc}
\hline
Notation& 
Description&
Notation& 
Description\\
\hline
$\mathcal{D}$&
Domain&
$s, t$&
Source, target\\
$X$&
Data set&
$Y$&
Label set\\
$x$&
Single sample&
$y$&
Corresponding label\\
$M$&
Source class number&
$N$&
Target class number\\
\multirow{2}{*}{$m$}&
Source samples number &
\multirow{2}{*}{$n$}&
Target samples number\\
&per class&&per class\\
$c$&
Class prototype&
$C$&
Prototype set\\
$P$&
Data distribution&
$f=h \circ g$&
Prediction function\\
$h$&
Feature extraction function&
$g$&
Classification function\\
\hline

\end{tabular}
\label{table7}
\end{table}

In this section, we describe the problem formulation and the proposed framework.
A source dataset $\mathcal{D}_s=\{(x_1^{si}, y_1^{si}),\dots,(x_j^{si}, y_j^{si}), \dots, (x_m^{si}, y_m^{si})\}_{i=1}^M$ is a collection of pairs $(x_j, y_j)$ where each $x_j\in \mathcal{X}^d$ is the $d-$dimensional feature vector and $y_j\in \mathcal{Y}$ is the corresponding label.  A target dataset is denoted by $\mathcal{D}_t=\{(x_1^{ti}, y_1^{ti}),\dots,(x_j^{ti}, y_j^{ti}), \dots, (x_n^{ti}, y_n^{ti})\}_{i=1}^N$. The $M,N$ denote the number of classes in datasets, $m,n$ denote the number of samples used in training of each class and $X^s,X^t$ denote the sets of $x_j^s,x_j^t$ respectively. 
The Table \ref{table7} describes the notations which are used in this work frequently. 

We assume that the probability distributions of  $X^s,X^t$  are different, i.e. $P(X^s)\neq P(X^t)$ and learn a prediction function $f: \mathcal{X}\to\mathcal{Y}$ which can work well on the target dataset. In general, $f$ is composed of two functions, $f(x,\theta)=g(h(x, \theta_1), \theta_2)$. Here $h:\mathcal{X}\to\mathcal{Z}$, the feature extraction function, is a mapping from the input space $\mathcal{X}$ to a latent space $\mathcal{Z}$, and $g:\mathcal{Z}\to\mathcal{Y}$ is a classification function to predict the input label. The $x$ and $\theta$ denote the  input and parameters of the model respectively. 
In our framework, we use a convolutional neural network as the feature extractor $h$, and learn several prototypes on the extracted features for each class to predict the corresponding label. In order to simplify the model, we only learn one prototype for each class in this work, and the prototype is denoted as $c_i$ where $i\in \{1,2,\dots,N\}$ represents the index of  predicted class prototype. The $c_i\in C^p$ is the p-dimensional vector.
\subsection{Architecture}
We selected the one-dimensional convolutional neural networks with wide first-layer kernels \cite{zhang2017new} for the framework. The model architecture contains two parts, one is the feature extractor, and the other is the classification layer. The input of the feature extractor is a multi-dimensional time series. The convolutional layers perform some non-linearities to convert it into  high-dimensional features.
Therefore, the feature extractor's output is an abstract representation of the input. Subsequently, we employed dropout \cite{hinton2012improving} on the feature extractor's output layer  for regularization. Dropout  randomly zeros some hidden units with a rate during training.
Finally, we used the prototypical layer instead of the traditional classification layers. The prototypical layer transforms  the high dimensional features into a p-dimensional vector which is used to   approximate the corresponding prototype.

This framework is based on the assumption that there exists a latent space that could minimize the discrepancy of the  same classes of source and target domains despite the similarity of  their samples. It means that the model can work well when  each class has its unique features, even if  the distribution of the class in the source domain is different from  the same class in the target domain. Meanwhile, we can learn  prototypes to represent each class in this latent space by the neural networks. Compared with traditional classification layers, the sample features of the same class can make a certain degree of change around the prototypes, which improves the generalization performance of the model.

\subsection{Feature extractor}

With the covariate shift assumption of domain adaptation\cite{ben2010impossibility}, we could get the assumption that $P(Y^s|Z^s)=P(Y^t|Z^t)$ 
when we learn a domain invariant space for the source and target distributions.
It means that the source and target domain classifiers  could be the same, i.e., $g_s=g_t$. Meanwhile, the parameters of CNN can be shared in a Siamese architecture, i.e., $h_s=h_t$. Therefore, we can directly apply the model learned on training data to the target dataset.

In that case, the method assumes that $h_s=h_t$, and we could train the function $h$ by minimizing a distance loss
\begin{equation}
\mathcal{L}_{D}(h)=E\left[\ell\left(d\left(h\left(x^{s}\right), h\left(x^{t}\right)\right),y_d\right)\right]
\label{eq1}
\end{equation}
where the $E[\cdot]$ denotes the mathematical  expectation and the $Y_d\in\{0,1\}$ denotes the consistency of two input streams.
One of the input streams comes from the source domain, and the other comes from the target domain.
When the two input streams are from the same classes, $Y_d=1$, otherwise, $Y_d=0$. Therefore, we take Binary Cross-Entropy Loss   as $\ell$, and the function $d$ could be any metrics for similarity measurement.
In this work, the function $d$ is computed with   Euclidean distance, $d\left(h\left(x^{s}\right), h\left(x^{t}\right)\right) = \sigma\left(\gamma\left\|h(x^s)-h(x^t)\right\|_2\right)$, where  $\sigma$ denotes the Sigmoid function and $\gamma$ is a hyper-parameter that controls the distance mapping.
The purpose of (\ref{eq1}) is to minimize the discrepancy between the source and target features in the latent space. 

The graphic description of our model can be found in Fig. \ref{fig1}.
This model uses the Siamese architecture based on  deep convolutional neural networks as the feature extractor. The CNN architecture is detailed in Table \ref{tab1}. In order to avoid the interference of  the high-frequency  noises in industrial environments, the model uses  wide kernels to extract features in the first layer and then uses small kernels to get better feature representations.

\begin{table}[t]
\renewcommand{\arraystretch}{1.3}
\caption{Structure of the feature extractor}
\centering
\setlength{\tabcolsep}{3pt}
\begin{tabular}{c|c|c}
\hline
Layer& 
Name&
Size/Stride\\
\hline
1& 
Convolutional-ReLU&
16 filters of $64\times1/1\times1$
\\

\hline
2& 
Max-Pooling&
$2\times1/2\times1$
\\

\hline
3& 
Convolutional-ReLU&
32 filters of $3\times1/1\times1$
\\

\hline
4& 
Max-Pooling&
$2\times1/2\times1$
\\

\hline
5& 
Convolutional-ReLU&
64 filters of $2\times1/1\times1$
\\

\hline
6& 
Max-Pooling&
$2\times1/2\times1$
\\

\hline
7& 
Convolutional-ReLU&
64 filters of $3\times1/1\times1$
\\

\hline
8& 
Max-Pooling&
$2\times1/2\times1$
\\

\hline
9& 
Convolutional-ReLU&
64 filters of $3\times1/1\times1$
\\

\hline
10& 
Max-Pooling&
$2\times1/2\times1$
\\

\hline
11& 
Fully-connected-Sigmoid&
100\\
\hline
\end{tabular}
\label{tab1}
\end{table}

\subsection{Classification layer}
With the abstract representations extracted by the feature extractor, traditional methods usually  use loss functions (for instance, Categorical Cross-Entropy for multi-class classification (\ref{eq3})) to minimize classification loss.
\begin{equation}
\mathcal{L}_{CA}(g)=-\sum_{i=1}^{N} y_{i} \log \left(g_{i}\right)
\label{eq3}
\end{equation}
The $y_i$ and $g_i$ are groundtruth and the model prediction for each class $i$ in $N$.

\begin{table}[t]
\renewcommand{\arraystretch}{1.3}
\caption{Structure of the classification layer}
\centering
\setlength{\tabcolsep}{3pt}
\begin{tabular}{c|c|c}
\hline
\multicolumn{3}{c}{Structure of the prototypical layer}\\
\hline
Layer& 
Name&
Parameter\\
\hline

12& 
Dropout&
$\text{rate}=0.5$\\
\hline
13& 
Fully-connected&
$\text{p}=5$\\
\hline
\multicolumn{3}{c}{Structure of the traditional layer}\\
\hline
Layer& 
Name&
Parameter\\
\hline

12& 
Dropout&
$\text{rate}=0.5$\\
\hline
13& 
Fully-connected-Softmax&
$\text{dimension}=N$\\
\hline

\end{tabular}
\label{tab7}
\end{table}

In this work, we combine the feature extractor with a prototypical layer to modify the model. In order to simplify the model, we only learn one prototype for each class, and the prototypical layer computes a p-dimensional vector  to  approximate the corresponding prototype.
After the prototypical layer described in Table \ref{tab7}, the abstract representations are converted to p-dimensional vectors. 
To learn the prototypes, a distance metric is used to compute the similarity between the p-dimensional vectors and the p-dimensional prototypes. Therefore, the model has two parts of trainable parameters, one for the prediction function $f$ and the other for the prototypes $C^p$. The distance can be measured by the function, e.g., $d(g(h(x)),c_i) =\|g(h(x))-c_i\|_2^2$. Finally, we define the classification loss as
\begin{equation}
\mathcal{L}_{CB}(g)=\mathcal{L}_{C}(g)+\lambda_1\|g-c_m\|_1-
\lambda_2\sum_{i\ne j}\|c_i-c_j\|_1+\lambda_3\sum_{i}^{N}\|c_i\|_2
\label{eq4}
\end{equation}
where the $\mathcal{L}_C(g)$  controls the classification accuracy,  $c_m$ is the corresponding prototype  of input $x$ and the 
$\lambda_i$ ($i\in \{1, 2, 3\}$) is the hyper-parameter that changes the weight of regularizations. The loss $\mathcal{L}_{C}(g)$ is defined as
\begin{alignat}{2}
\mathcal{L}_{C}(g)&=E\left[\ell\left(ds\left(g\left(h(x)\right), C\right),y\right)\right]\\
ds\left(g\left(h(x)\right), c_i\right)&= \frac{e^{-\gamma d\left(g(h(x)), c_{i}\right)}}{\sum_{l=1}^{N} e^{-\gamma d\left(g(h(x)), c_{l}\right)}}
\end{alignat}
where $\gamma$ is a hyper-parameter that can change the hardness of probability assignment\cite{yang2018robust}.
By limiting the distance between  vectors and the corresponding prototypes, the regularization $\|g-c_m\|_1$ makes the features in the same classes more compact. Meanwhile, the regularization
$\sum_{i\ne j}\|c_i-c_j\|_1+\sum_{i}^{N}\|c_i\|_2$ makes the different classes separated. By combining these regularizations, we can make the model more robust and improve  classification accuracy. 

Therefore, we get the modified approach by learning a deep model $f$ such that
\begin{equation}
\mathcal{L}_{P}(f)=\lambda\mathcal{L}_{D}(h)+(1-\lambda)\mathcal{L}_{CB}(g)
\label{eq7}
\end{equation}
The prototypical network $g$ is trained with source data, and then fine-tuned based on the  few samples in $\mathcal{D}_t$
\begin{equation}
g_t = \text{fine-tune}(h|\mathcal{D}_t)
\label{eq8}
\end{equation}
Finally, We use minibatch stochastic gradient descent and AdaDelta\cite{zeiler2012adadelta} with hyper-parameters $\epsilon=10^{-6}$ and $\rho=0.9$
to train the model $f$ and prototypes $C^p$. The model is initialized by the weight initialization in \cite{he2015delving} and trained with a minibatch size of 64 on a single GPU.

\section{Experiments}

\begin{table*}
\renewcommand{\arraystretch}{1.3}
\caption{Description of the datasets}
\centering
\setlength{\tabcolsep}{3pt}
\begin{tabular}{c|c|c|c|c|c|c|c|c|c|c|c|c|c}
\hline
\textbf{Drive-end Data} & Normal & \multicolumn{3}{|c|}{Rolling Element} &\multicolumn{3}{|c|}{Inner Race}  &\multicolumn{3}{|c|}{Outer Race (6
o'clock)} & Speed (rpm)&Description&Number of samples\\
\hline
Fault Diameter (inch) & 0 & 0.007&0.014&0.021& 0.007&0.014&0.021& 0.007&0.014&0.021&\multirow{2}{*}{1730}&\multirow{2}{*}{Dataset A}&\multirow{2}{*}{1250$\times$10}\\
\cline{1-11}
Fault Labels &0&1&2&3&4&5&6&7&8&9&&\\
\hline

\textbf{Fan-end Data}& Normal & \multicolumn{3}{|c|}{Rolling Element} &\multicolumn{3}{|c|}{Inner Race}  &\multicolumn{3}{|c|}{Outer Race(6
o'clock)} & Speed (rpm)&Description&Number of samples\\
\hline
Fault Diameter (inch) & 0 & 0.007&0.014&0.021& 0.007&0.014&0.021& 0.007&0.014&0.021&\multirow{2}{*}{1797}&\multirow{2}{*}{Dataset B}&\multirow{2}{*}{1250$\times$10}\\
\cline{1-11}
Fault Labels &0&1&2&3&4&5&6&7&8&9&&\\
\hline

\textbf{Drive-end Data} & Normal & \multicolumn{3}{|c|}{Rolling Element} &\multicolumn{3}{|c|}{Inner Race}  &\multicolumn{3}{|c|}{Outer Race(6
o'clock)} & Speed (rpm)&Description&Number of samples\\
\hline
Fault Diameter (inch) & 0 & 0.007&0.014&0.021& 0.007&0.014&0.021& 0.007&0.014&0.021&\multirow{2}{*}{1730}&\multirow{2}{*}{Dataset C}&\multirow{2}{*}{1250$\times$6}\\
\cline{1-11}
Fault Labels &0&1&2&*&4&*&*&7&*&9&&\\
\hline

\textbf{Fan-end Data} & Normal & \multicolumn{3}{|c|}{Rolling Element} &\multicolumn{3}{|c|}{Inner Race}  &\multicolumn{3}{|c|}{Outer Race(6
o'clock)} & Speed (rpm)&Description&Number of samples\\
\hline
Fault Diameter (inch) & 0 & 0.007&0.014&0.021& 0.007&0.014&0.021& 0.007&0.014&0.021&\multirow{2}{*}{1797}&\multirow{2}{*}{Dataset D}&\multirow{2}{*}{1250$\times$6}\\
\cline{1-11}
Fault Labels &0&1&2&*&4&*&*&7&*&9&&\\
\hline

\textbf{Fan-end Data} & Normal & \multicolumn{3}{|c|}{Rolling Element} &\multicolumn{3}{|c|}{Inner Race}  &\multicolumn{3}{|c|}{Outer Race(6
o'clock)} & Speed (rpm)&Description&Number of samples\\
\hline
Fault Diameter (inch) & 0 & 0.007&0.014&0.021& 0.007&0.014&0.021& 0.007&0.014&0.021&\multirow{2}{*}{1797}&\multirow{2}{*}{Dataset E}&\multirow{2}{*}{1250$\times$10}\\
\cline{1-11}
Fault Labels &0&9&6&3&2&5&7&8&4&1&&\\
\hline
\end{tabular}
\label{tab2}
\end{table*}
\subsection{Data description}
\begin{figure}
\centerline{\includegraphics[width=3in]{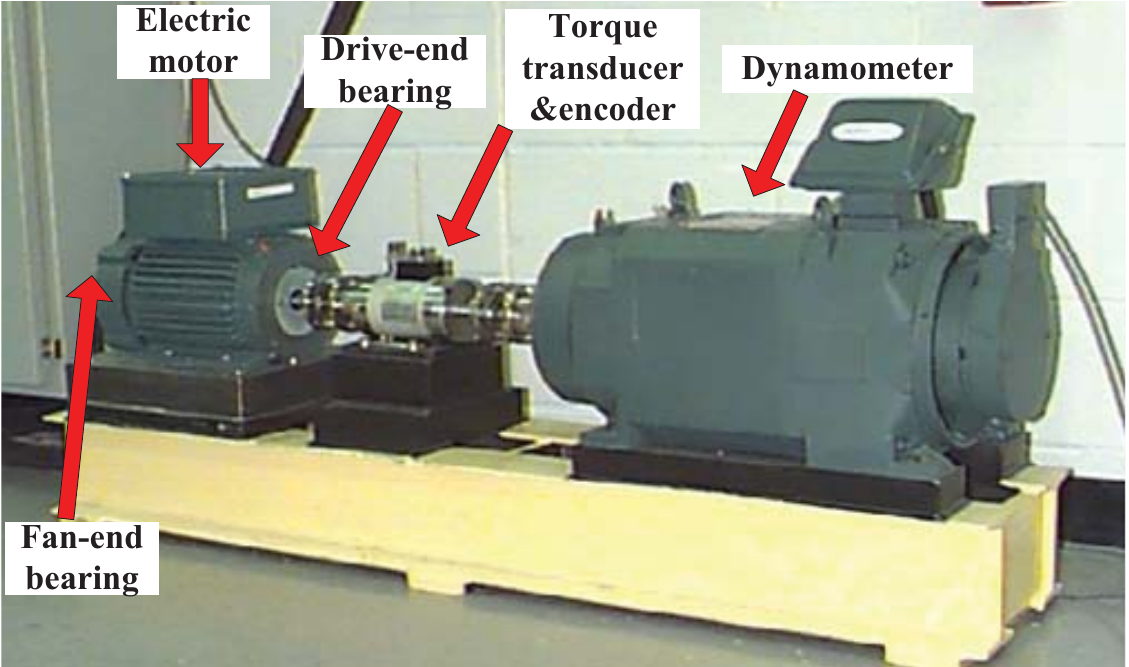}}
\caption{CWRU bearing test stand\cite{article}.
}
\label{fig5}
\end{figure}

We used the  Case Western Reserve University Bearing Datasets\cite{CWRU} to evaluate the performance of our framework. 
The layout of the test stand is shown in Fig. \ref{fig5}.
Data were collected in three situations for normal bearings, single-point drive-end and fan-end defects with diameter from 0.007 to 0.028 (SKF bearings) inches. 
There are three types of data according to the  location of defects: inner race fault,  rolling element fault, and outer race fault. The outer race fault has three categories  based on the fault position relative to the load zone:  at 3 o'clock, at 6 o'clock, and at 12 o'clock.
Each type was recorded for motor loads of 0 to 3 horsepower (speeds of 1797 to 1720 rpm) at 12kHz (some samples at 48kHz). A detailed description can be found in \cite{article}.
\begin{figure}
\centerline{\includegraphics[width=3.5in]{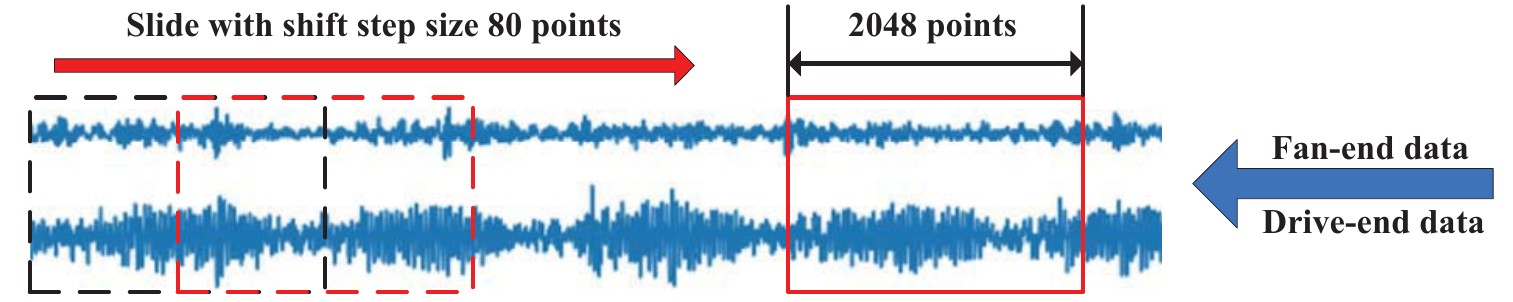}}
\caption{Data generation with overlap.
}
\label{fig2}
\end{figure}

In this work, we selected 12kHz drive-end and fan-end bearing fault data to verify the adaptation ability between the datasets from different ends.  As shown in Fig. \ref{fig2}, the samples are generated by the sliding window of 2048 size with 80 points shift step. 
In the experiments, we selected the fault diameters of 0.007, 0.014, and 0.021 inches  for every type of fault and had ten conditions in total added with a normal condition. Datasets A, B, C, D, and E each contain 1250 samples per class, and Datasets C and D have only six classes. The details of all the datasets are described in Table \ref{tab2}.

In order to verify the effectiveness of the framework, we trained models by these methods: \textbf{SVM}, \textbf{CTM} (the feature extractor + traditional classification layer ($\mathcal{L}_{CA}(f)$)), \textbf{FTM} (the feature extractor ($\mathcal{L}_{D}(h))$ + traditional classification layer ($\mathcal{L}_{CA}(f)$)) and \textbf{FPM} (the feature extractor ($\mathcal{L}_{D}(h)$) + prototypical layer ($\mathcal{L}_{CB}(f)$)). The SVM and CTM are trained by  using the source data and n samples per class in the target dataset without domain adaptation. We randomly selected n samples per class in the target dataset for four times to generate different training sets and repeated the training for five times to calculate the mean of accuracy.
In  subsection E, we compared our  framework with the \textbf{WDMAN} proposed in \cite{zhang2019deep}. Our framework achieved  competitive results with  only one target sample per class in training.

\begin{table}
\renewcommand{\arraystretch}{1.3}
\caption{Test accuracy(\%) of tasks A$\leftrightarrows$B}
\centering
\setlength{\tabcolsep}{4pt}
\begin{tabular}{c|c|c|c|c|c|c}
\hline
Task  (\%)& \multicolumn{3}{|c|}{A$\to$B} &\multicolumn{3}{|c}{B$\to$A}\\
\hline
n  samples (per class) &1&2&3&1&2&3\\
\hline
SVM &\multicolumn{3}{|c|}{23.09}&\multicolumn{3}{|c}{22.51}\\
\hline
CTM &48.83&58.32&67.48&47.84&64.71&69.74\\
\hline
FTM &86.28&98.22&99.46&95.43&99.80&99.91\\
\hline
FPM &92.91 &98.52&99.55&97.53&99.56&99.72 \\
\hline
\end{tabular}
\label{tab3}
\end{table}


\begin{figure}
\centering
\subfigure[Source domain with FPM]{
\includegraphics[width=3.9cm]{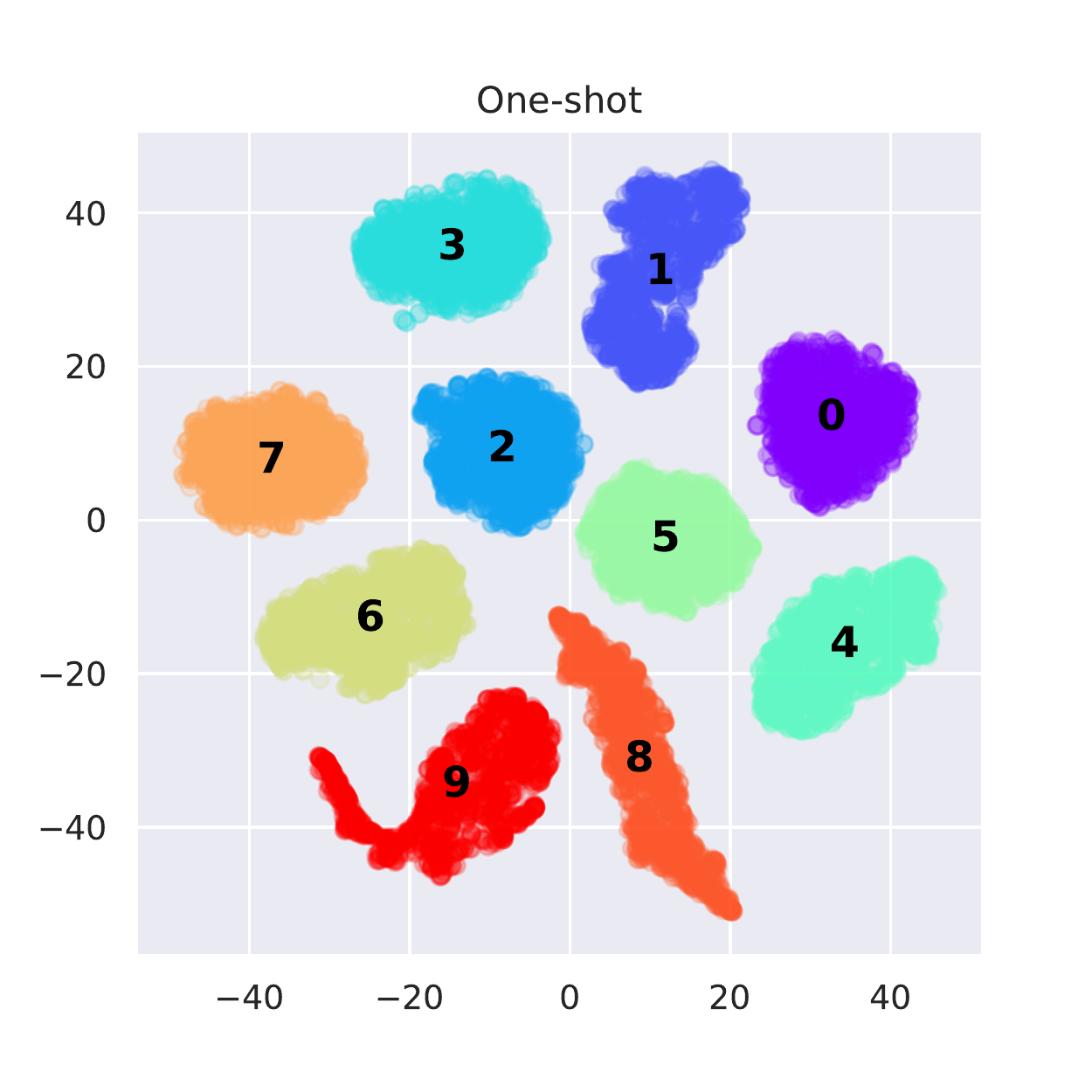}
}
\quad
\subfigure[Source domain with FPM]{
\includegraphics[width=3.9cm]{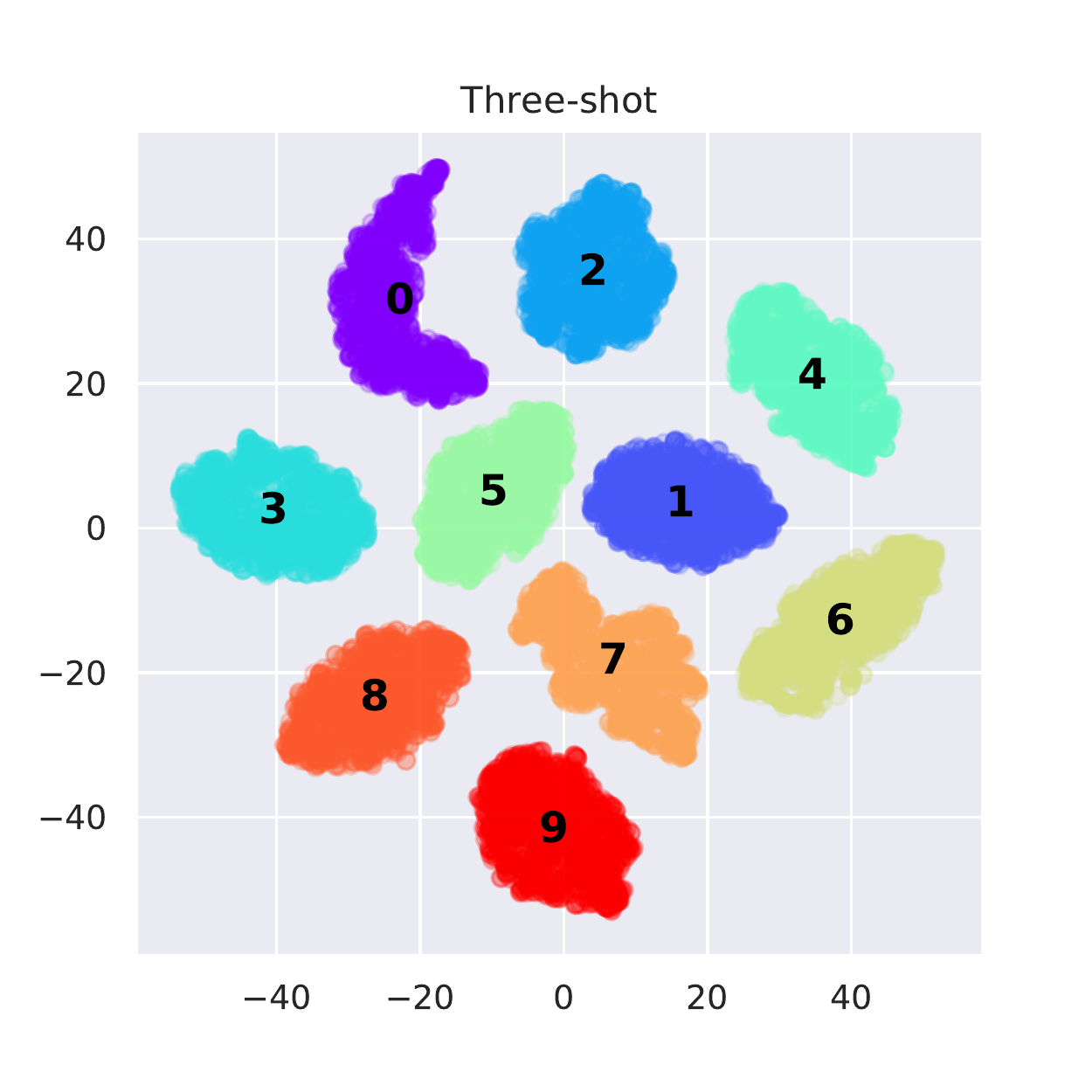}
}
\quad
\subfigure[Target domain with FPM]{
\includegraphics[width=3.9cm]{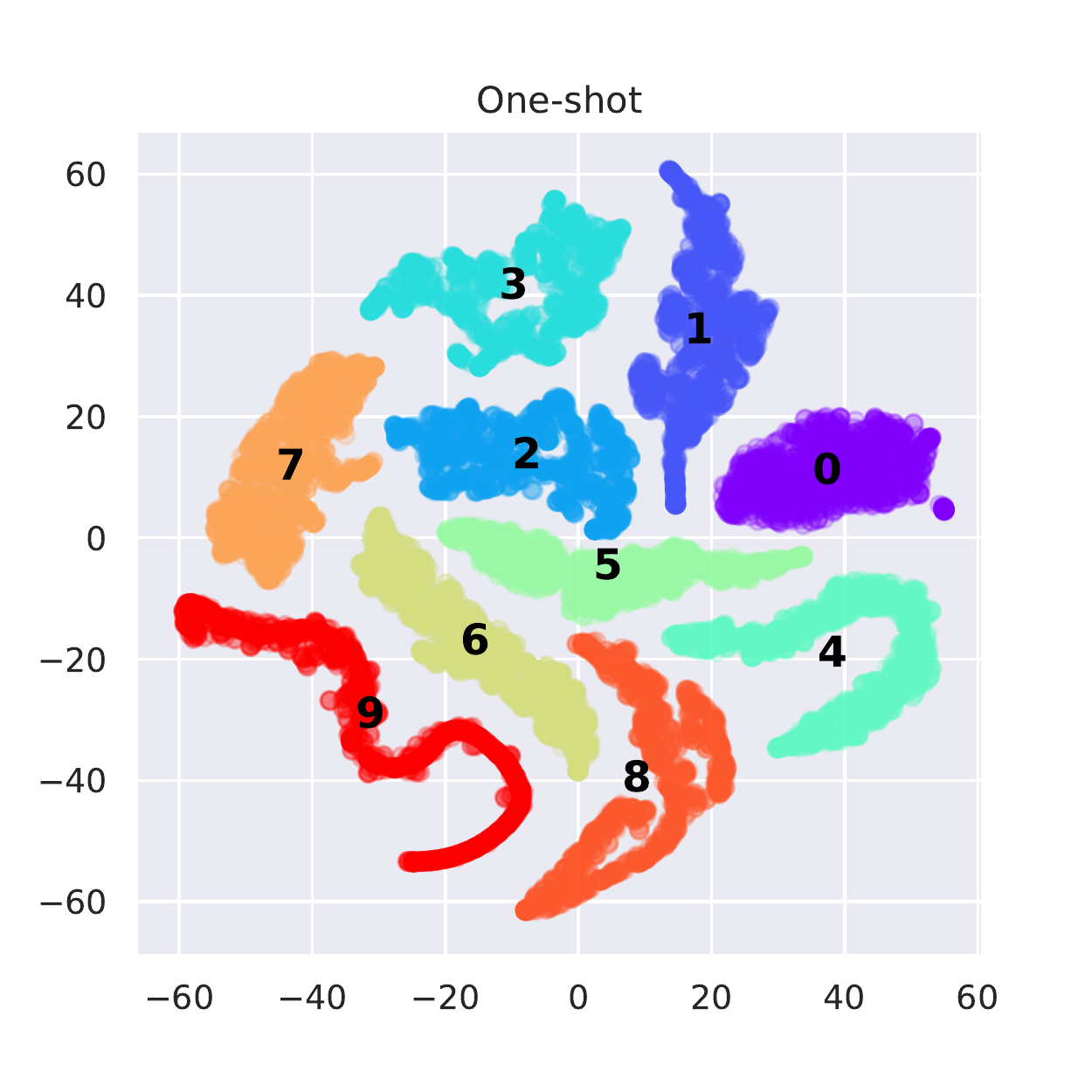}
}
\quad
\subfigure[Target domain with FPM]{
\includegraphics[width=3.9cm]{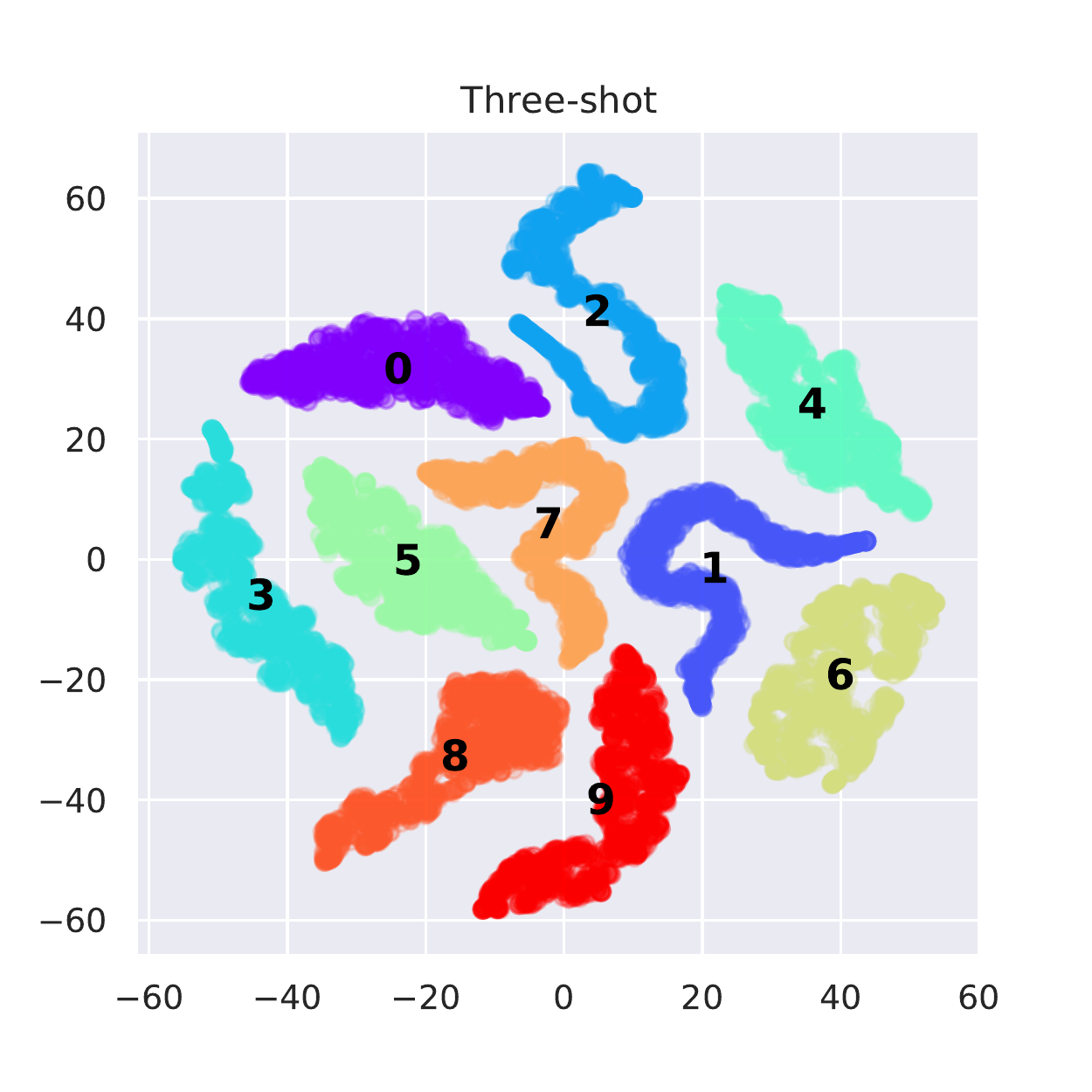}
}
\caption{Feature visualization of task A$\to$B:
100-dimensional vectors learned by the feature extractor are reduced to a two-dimensional map.}
\label{fig3}
\end{figure}

\subsection{Few-shot  domain adaptation with complete classes in the source domain}
In the first experiment, we conducted experiments on the dataset A and  B to evaluate the  adaptation ability of the proposed framework. 
We randomly selected n ($n\in\{1,2,3\}$) samples per class in the target dataset and utilized the  source dataset that contains 1250 samples per class for training.
Moreover, the rest of the target data are used for testing.  

Table \ref{tab3} reports  that the classification accuracy of four methods and the number of target training samples has little effect on the performance of SVM. The proposed framework can work well even when we use only ten labeled target samples (n=1, one sample per class) for training. We also trained the base model by CTM to get the lower bound using source and a few target samples without domain adaptation.
To demonstrate the adaptation ability directly, we followed the \textit{t-SNE}\cite{maaten2008visualizing} to  visualize the high dimensional features in a two-dimensional map.  
Fig. \ref{fig3} shows the visualizations of the source and target reduced features. 
As shown in Fig. \ref{fig3}, the target features learned by FPM  have visible  class prototypes that are close to the centers of source features, which shows that the FPM can make better use of inter-class information.
Moreover, the target features  are inter-class separable with only one sample per class. However, there is a problem that some samples in one class are closer to the prototypes of other classes. When using three samples per class,  the target features within the same class are more compact and distinguish.


\subsection{Few-shot  domain adaptation with incomplete classes in the source domain}

\begin{table}
\renewcommand{\arraystretch}{1.3}
\caption{Test accuracy(\%) of tasks C$\to$B and D$\to$A}
\centering
\setlength{\tabcolsep}{4pt}
\begin{tabular}{c|c|c|c|c|c|c}
\hline
Task  (\%)& \multicolumn{3}{|c|}{C$\to$B} &\multicolumn{3}{|c}{D$\to$A}\\
\hline
n samples (per class) &1&3&5&1&3&5\\
\hline
SVM &\multicolumn{3}{|c|}{16.68}&\multicolumn{3}{|c}{30.14}\\
\hline
CTM &46.97&47.94&55.68&52.16&53.25&59.94\\
\hline
FTM &58.05&78.73&92.61&63.18&81.52&98.71\\
\hline
FPM &63.24&85.67&95.33&66.87&86.95&98.63 \\
\hline
\end{tabular}
\label{tab4}
\end{table}

\begin{figure}
\centering
\subfigure[Source domain with FPM]{
\includegraphics[width=3.9cm]{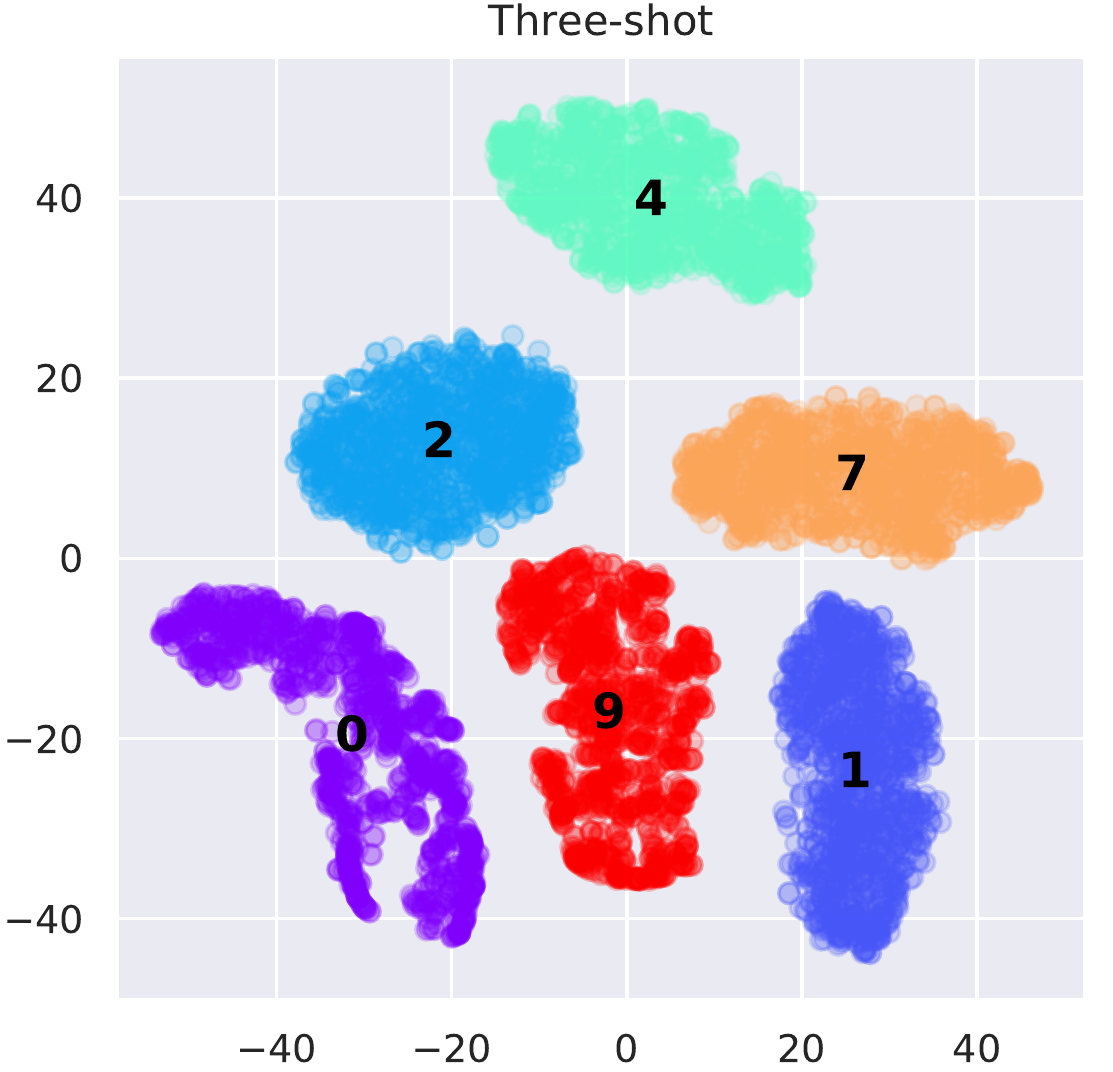}
}
\quad
\subfigure[Source domain with FPM]{
\includegraphics[width=3.9cm]{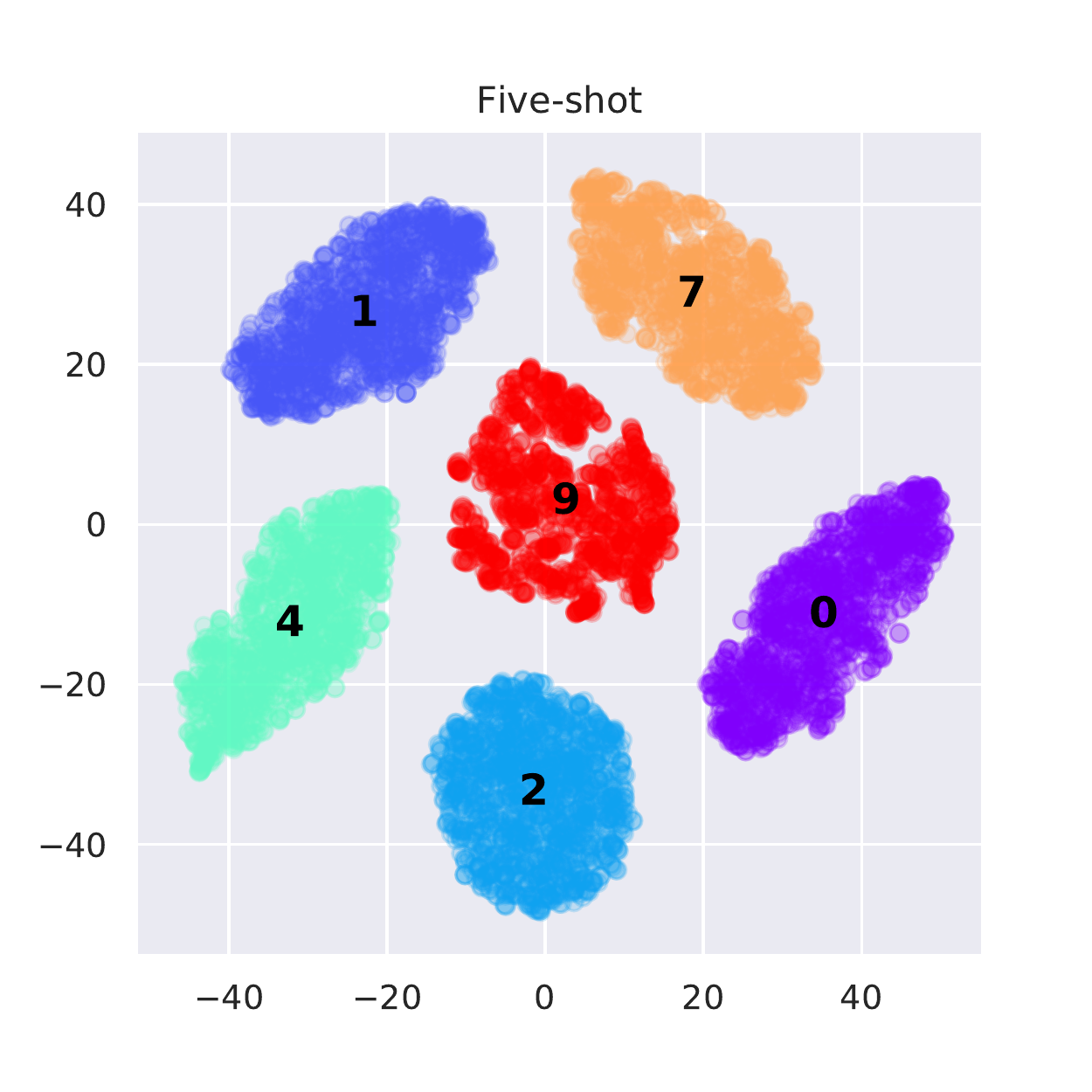}
}
\quad
\subfigure[Target domain with FPM]{
\includegraphics[width=3.9cm]{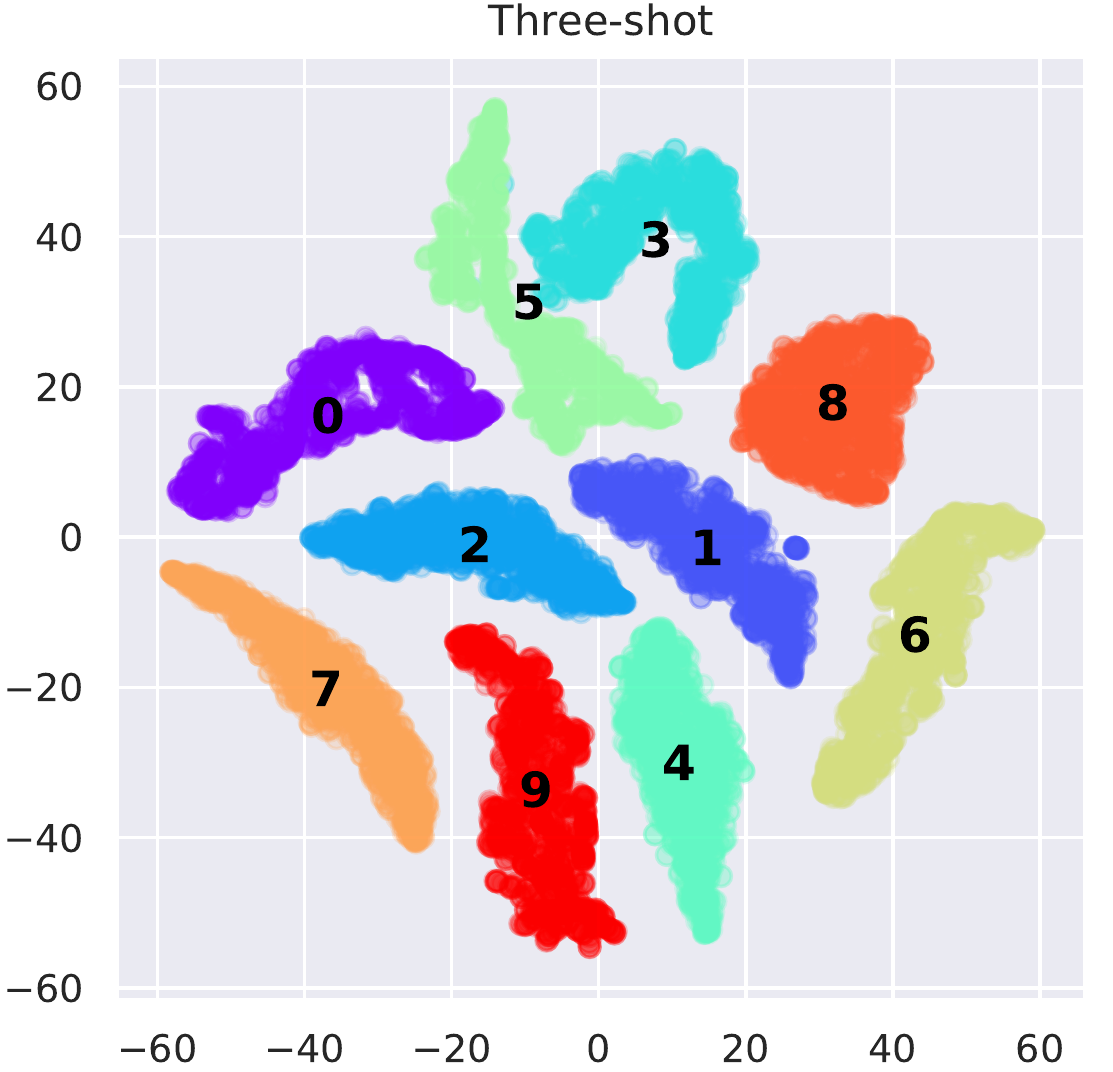}
}
\quad
\subfigure[Target domain with FPM]{
\includegraphics[width=3.9cm]{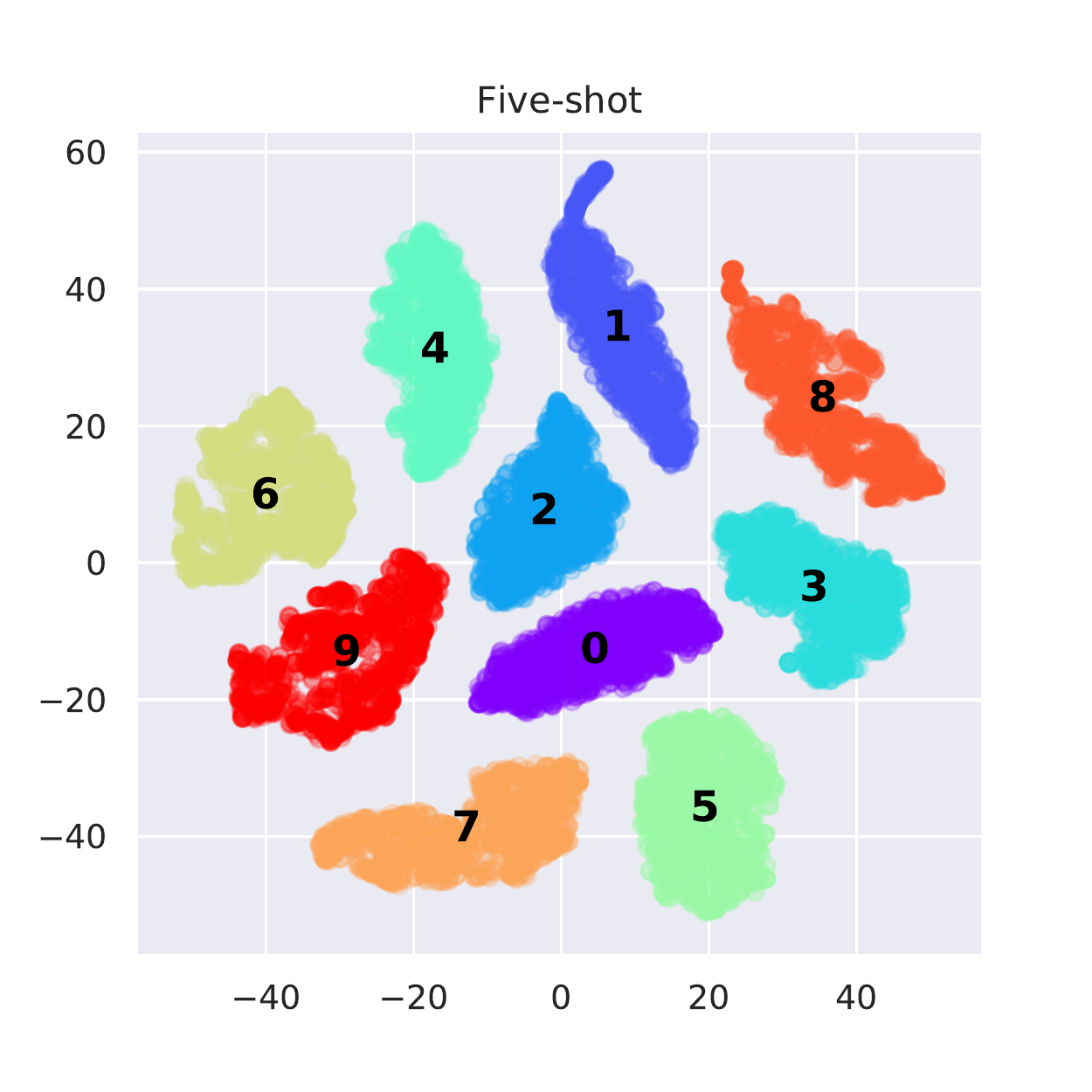}
}

\caption{Feature visualization of task D$\to$A:
100-dimensional vectors learned by the feature extractor are reduced to a two-dimensional map.}
\label{fig4}
\end{figure}

\begin{figure}
\centering
\subfigure[Confusion matrix with FPM]{
\includegraphics[width=2.6in]{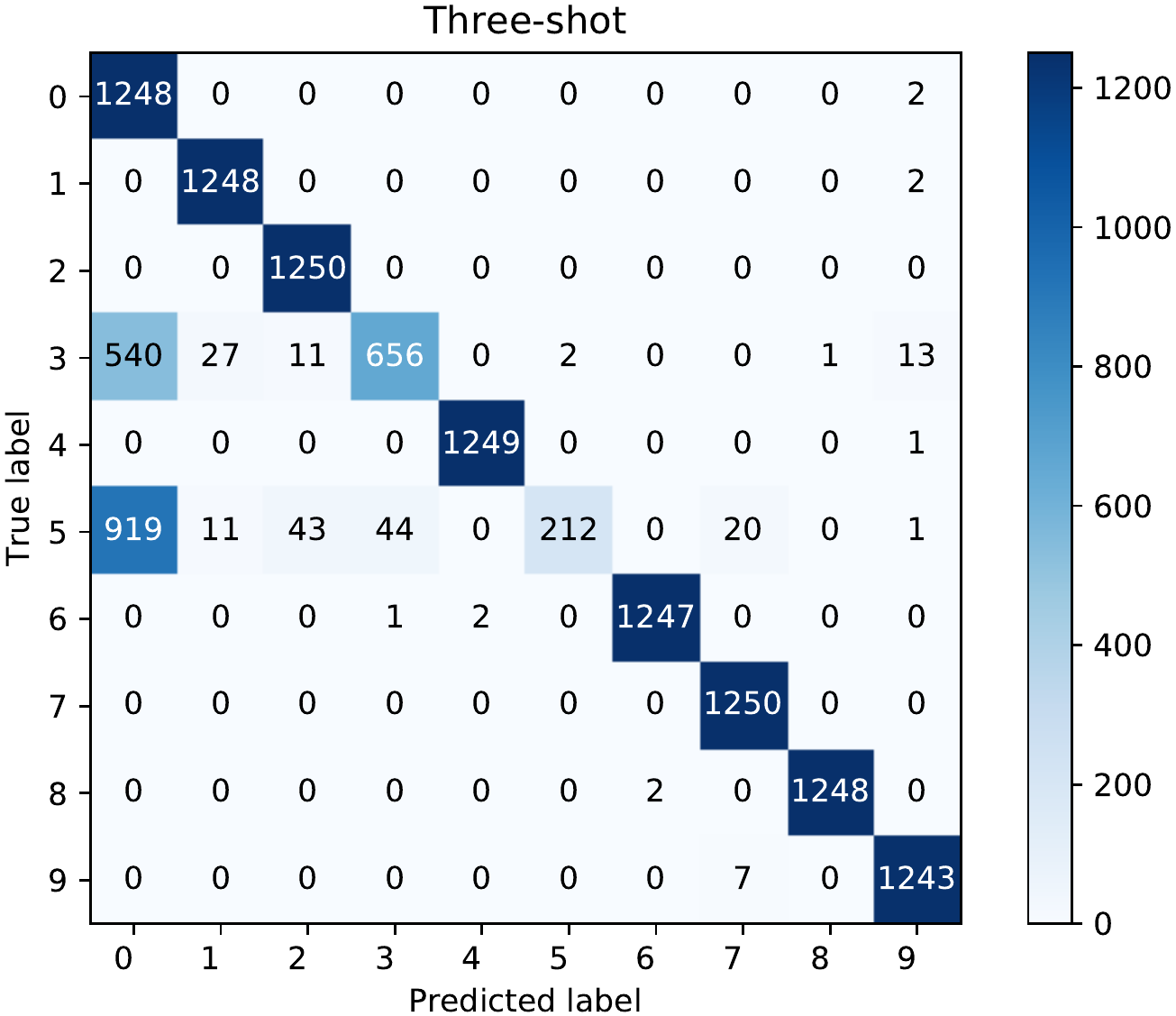}
}
\quad
\subfigure[Confusion matrix with FPM]{
\includegraphics[width=2.6in]{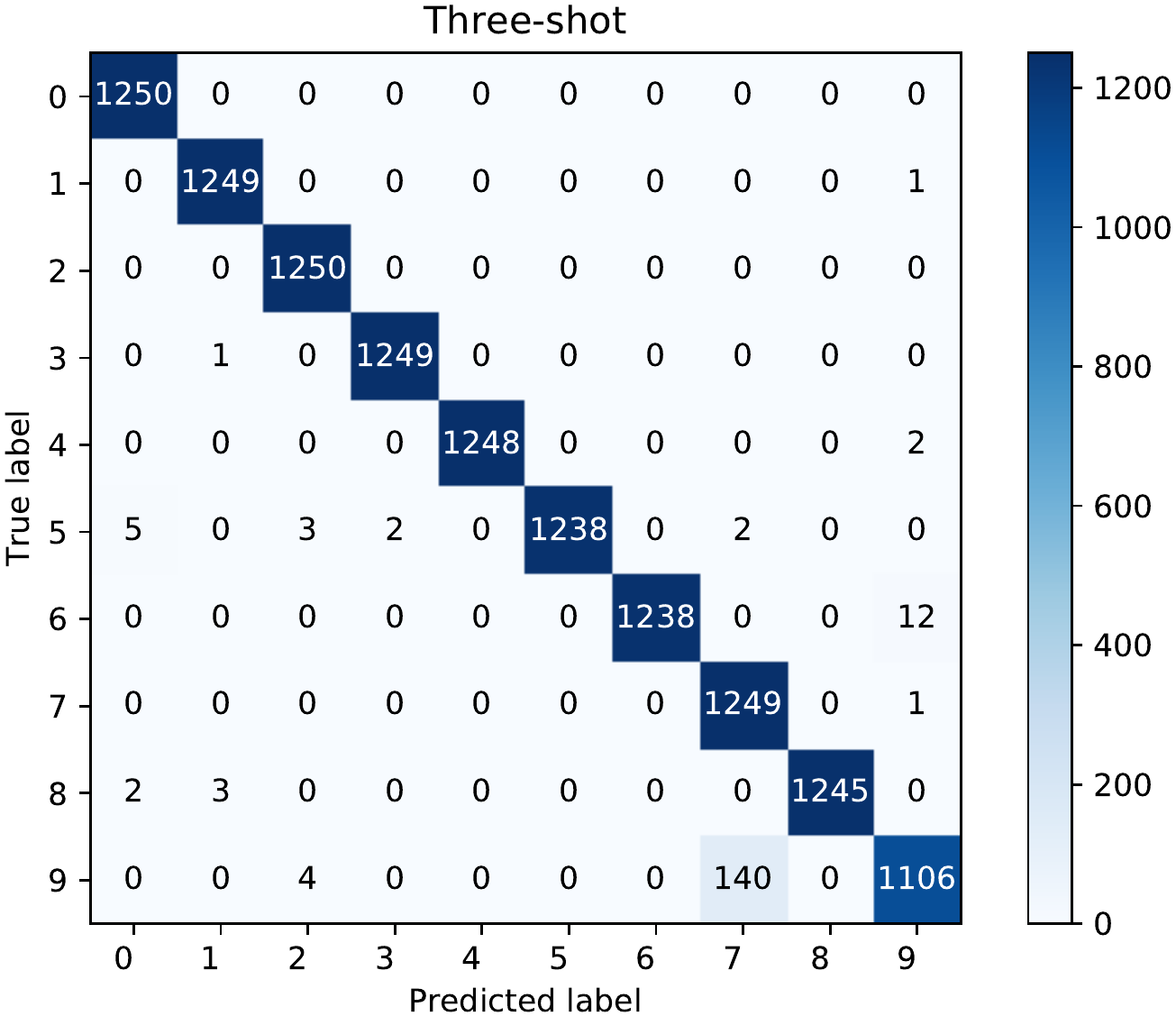}
}
\caption{Confusion matrix of task D$\to$A.}
\label{fig6}
\end{figure}

In real-world industrial applications, we could not obtain complete types of fault data. Therefore, we extend the model to address the problem when the classes from the source and target domains are not completely overlapping.
The model must generalization to unseen classes in the source domain,  given only a few examples of each new target class. In the second experiment, the model is adapted to accommodate ten classes in the  target dataset when we use  a few target samples and  the source dataset, which contains six classes in training.

Table \ref{tab4} shows the classification accuracies increasing with the number of target samples available in training ($n\in\{1,3,5\}$) rising, and FPM has improved accuracy compared with other methods.
Compared with the  traditional classification layer, the prototypical layer 
can learn the latent space where both seen and new classes are inter-class separable and intra-class compact, which proves that  features learned by FPM from raw signals are more domain invariant.
We visualized the features of task D$\to$A  in Fig. \ref{fig4}, which showed that the model learned by FPM is easier to distinguish  the new class features in the target domain.
As shown in the Fig. \ref{fig4}(c), the samples in class 3 are close to the prototype of class 5 when using three target samples per class in training. To better show the effect of FPM, we calculated the confusion matrix to visualize the classification results.
From Fig. \ref{fig6}(a), classes 3 and 5 perform poorly due to lack of training samples. 
The Fig. \ref{fig6}(b) shows that the model quickly converges with high classification accuracy as the number of target samples available in training increases.

\subsection{Few-shot  domain adaptation with randomized label assignments}

\begin{table}
\renewcommand{\arraystretch}{1.3}
\caption{Test accuracy(\%) of tasks  A$\leftrightarrows$E}
\centering
\setlength{\tabcolsep}{4pt}
\begin{tabular}{c|c|c|c|c|c|c}
\hline
Task  (\%)& \multicolumn{3}{|c|}{A$\to$E} &\multicolumn{3}{|c}{E$\to$A}\\
\hline
n samples (per class) &1&2&3&1&2&3\\
\hline
SVM &\multicolumn{3}{|c|}{19.16}&\multicolumn{3}{|c}{10.18}\\
\hline
CTM &43.86&45.27&62.43&52.98&57.85&70.22\\
\hline
FTM &81.71&97.62&98.82&92.42&98.73&99.40\\
\hline
FPM &92.30&99.12&99.65&95.86&99.46&99.73 \\
\hline
\end{tabular}
\label{tab5}
\end{table}

In real-world industrial applications, we  could not obtain any labeled target samples for some types of faults. With massive unlabeled target data, we can utilize the  most different samples to represent the respective types of faults.
Since we do not know the types of faults in advance, the representative target sample labels will be randomly scrambled. In order to verify the robustness of the proposed framework, we randomly disrupted  the class labels of  dataset B to get dataset E.
In the third experiment, we followed the setting of the first experiment but replaced the dataset B with dataset E.
From Table \ref{tab5}, there are little differences between the accuracies of the first task and third task, which proves the assumption in Section III.
The performance of the framework does not rely on the similarity between the same classes in source and target domains. It extracts unique features for each class and maps the same class features to a common latent space. Therefore, the proposed framework  can compensate for great domain shifts very well.

\subsection{Comparison experiments}
\begin{table}
\renewcommand{\arraystretch}{1.3}
\caption{Test accuracy(\%) of tasks described in \cite{zhang2019deep}}
\centering
\setlength{\tabcolsep}{5pt}
\begin{tabular}{c|c|c|c|c|c}
\hline
\multirow{2}{*}{Task (\%)} &WDMAN &\multicolumn{2}{|c|}{FTM}&\multicolumn{2}{|c}{FPM}\\
\cline{3-6}
 &\cite{zhang2019deep}&1&2&1&2\\
\hline
A$\to$B&99.73 &94.98     &99.12     &99.27 &99.73 \\
A$\to$C&99.67 &93.13     &97.51     &99.40 &99.94 \\
A$\to$D&100   &96.37     &99.79     &99.56 &99.90 \\
\hline
B$\to$A&99.13 &99.26     &99.81     &99.53 &99.84 \\
B$\to$C&100   &100       &100     &99.67 &99.93 \\
B$\to$D&99.93 &99.99     &100     &99.50 &99.91 \\
\hline
C$\to$A&98.53 &97.08     &98.46     &98.80 &99.25 \\
C$\to$B&99.80 &97.48     &99.35     &99.12 &99.76 \\
C$\to$D&100   &100       &100     &99.72 &99.95 \\
\hline
D$\to$A&98.07 &97.09     &98.35     &97.91 &98.13 \\
D$\to$B&98.27 &85.78     &94.33     &91.54 &99.03 \\
D$\to$C&99.53 &95.59     &99.12     &99.41 &99.78 \\
\hline

\end{tabular}
\label{tab6}
\end{table}

\begin{figure}
\centering
\subfigure[One-shot with FPM]{
\includegraphics[width=2.6in]{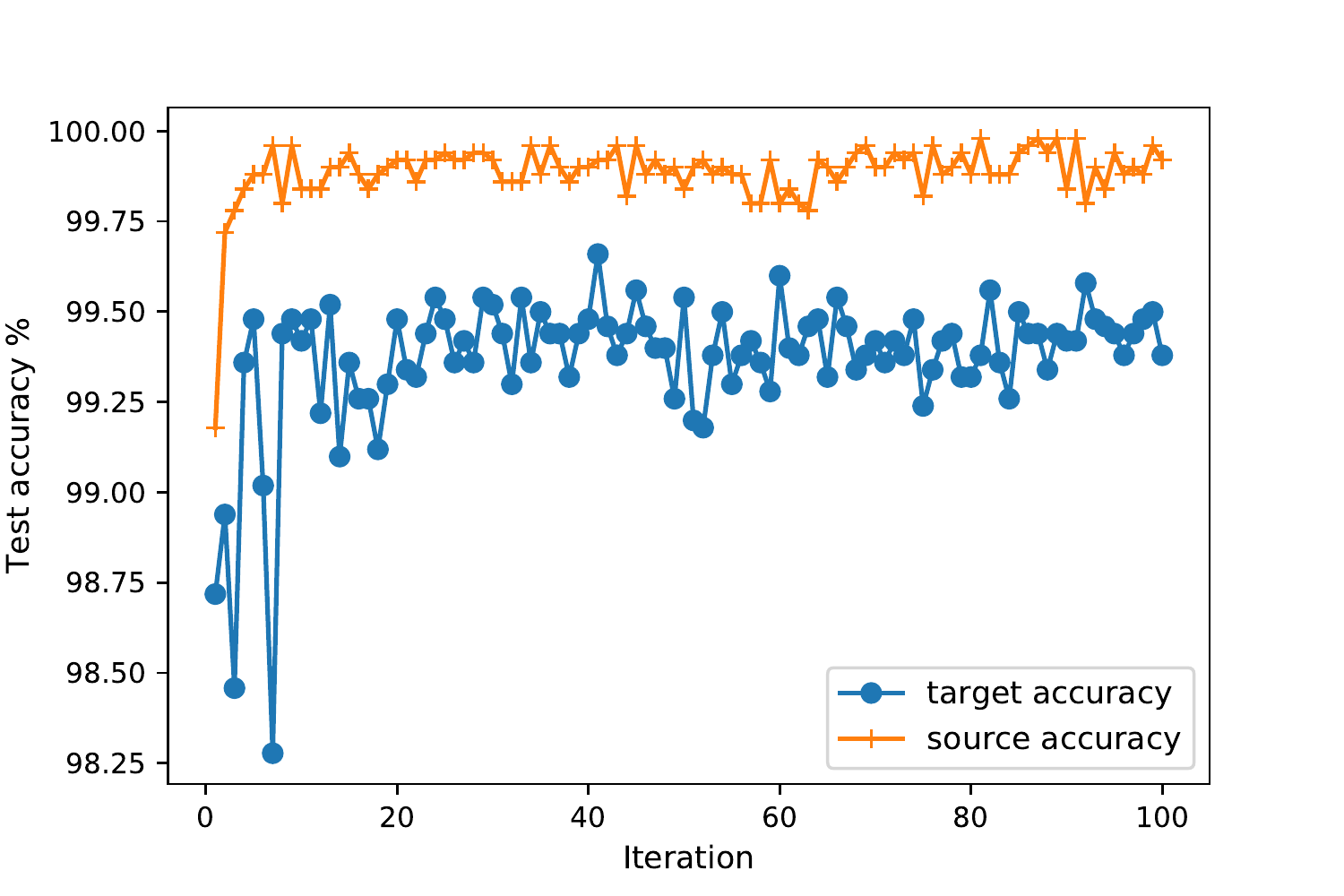}
}
\quad
\subfigure[Two-shot with FPM]{
\includegraphics[width=2.6in]{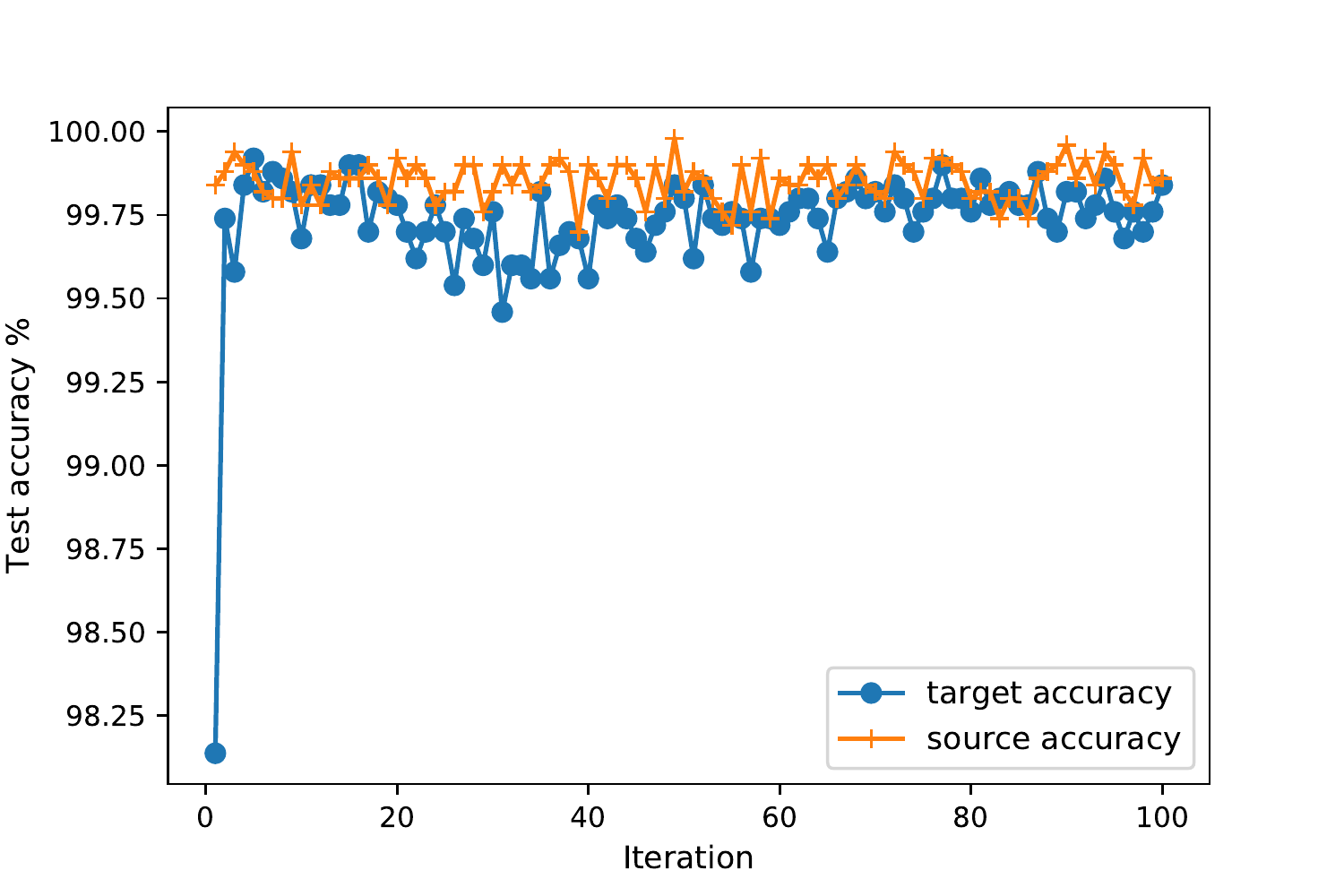}
}

\caption{Performance of FPM using different numbers of target samples.}
\label{fig7}
\end{figure}

We evaluated the performance of our framework on the CWRU bearing datasets and followed the experiment settings described in \cite{zhang2019deep}. 
Unlike the experiment settings above, the adaptation ability is verified on the data from the same end.
The  datasets A, B, C, and D were drawn from four different motor loads (0, 1, 2, and 3 hp) with a sampling frequency of 12 kHZ.
There are ten classes in total for each dataset, and their settings are the same as Table \ref{tab2}, which consists of nine types of fault and a normal condition.
Each class contains 500 samples whose sizes are equal to 1200.

As shown in Table \ref{tab6}, we compared our framework with the WDMAN proposed in \cite{zhang2019deep}.  With only one target sample per class in training, the FPM shows attractive results except for task D$\to$B. And we can improve the accuracy of task D$\to$B
to 99.03\% when using two target samples per class.
Although the WDMAN does not require the labeled target samples in training, our framework  requires very few labeled target samples, and the performance of the model can be improved as the labeled target samples increases. This framework is  simpler  than the unsupervised methods which use more than thousands of unlabeled target samples in training. Without the pre-training process, the model can be applied directly to the target dataset.  Compared with the arduous training of the Generative Adversarial Networks (GAN) in WDMAN, this model can quickly converge to a high accuracy as the  labeled target samples increase in training. To demonstrate the speed of convergence, we plotted the classification accuracy of task  A$\to$B in Fig. \ref{fig7}. As shown in Fig. \ref{fig7}, the target domain can achieve high accuracy with few training epochs by FPM. Compared with only one target sample per class in training, the classification accuracy of the target domain is very close to that of the source domain when using two samples, which shows that the framework has an effective adaptation ability. 

\section{Conclusion}
We have introduced a deep model in combination with domain adaptation and prototype learning for fault diagnosis.
This deep model takes raw temporal signals as inputs and achieves a high classification accuracy on the CWRU bearing datasets.
Without changing the model architecture,
the proposed framework can be applied to address the problem  when the classes from the source and target domains are not completely overlapping. 
Our experiments  show that the framework has an effective adaptation ability, which requires a few samples from a priori fixed target distribution. 
Moreover, the model accuracy  can converge quickly as the  labeled target samples increase in training. In future work, we will utilize other metrics  for similarity measurement (e.g., Maximum Mean Discrepancy) instead of Euclidean distance and increase the number of prototypes for each class. In the experiments, we found that there are some differences in model performance as the randomly selected target training data changes, especially when using only one sample per class in training.
For future work, we hope to further optimize the loss function  to reduce the number of hyper-parameters and improve  model stability. Overall, the effectiveness of the proposed framework makes it a promising method for fault diagnosis. 


%





\ifCLASSOPTIONcaptionsoff
  \newpage
\fi



%

\bibliographystyle{IEEEtran}
\bibliography{ff} 

\begin{thebibliography}{10}
\providecommand{\url}[1]{#1}
\csname url@samestyle\endcsname
\providecommand{\newblock}{\relax}
\providecommand{\bibinfo}[2]{#2}
\providecommand{\BIBentrySTDinterwordspacing}{\spaceskip=0pt\relax}
\providecommand{\BIBentryALTinterwordstretchfactor}{4}
\providecommand{\BIBentryALTinterwordspacing}{\spaceskip=\fontdimen2\font plus
\BIBentryALTinterwordstretchfactor\fontdimen3\font minus
  \fontdimen4\font\relax}
\providecommand{\BIBforeignlanguage}[2]{{%
\expandafter\ifx\csname l@#1\endcsname\relax
\typeout{** WARNING: IEEEtran.bst: No hyphenation pattern has been}%
\typeout{** loaded for the language `#1'. Using the pattern for}%
\typeout{** the default language instead.}%
\else
\language=\csname l@#1\endcsname
\fi
#2}}
\providecommand{\BIBdecl}{\relax}
\BIBdecl

\bibitem{li2000stochastic}
Y.~Li, T.~Kurfess, and S.~Liang, ``Stochastic prognostics for rolling element
  bearings,'' \emph{Mechanical Systems and Signal Processing}, vol.~14, no.~5,
  pp. 747--762, 2000.

\bibitem{dong2007hidden}
M.~Dong and D.~He, ``Hidden semi-markov model-based methodology for
  multi-sensor equipment health diagnosis and prognosis,'' \emph{European
  Journal of Operational Research}, vol. 178, no.~3, pp. 858--878, 2007.

\bibitem{simani2003model}
S.~Simani, C.~Fantuzzi, and R.~J. Patton, ``Model-based fault diagnosis
  techniques,'' in \emph{Model-based Fault Diagnosis in Dynamic Systems Using
  Identification Techniques}.\hskip 1em plus 0.5em minus 0.4em\relax Springer,
  2003, pp. 19--60.

\bibitem{lee2016convolutional}
D.~Lee, V.~Siu, R.~Cruz, and C.~Yetman, ``Convolutional neural net and bearing
  fault analysis,'' in \emph{Proceedings of the International Conference on
  Data Mining (DMIN)}.\hskip 1em plus 0.5em minus 0.4em\relax The Steering
  Committee of The World Congress in Computer Science, Computer~…, 2016, p.
  194.

\bibitem{frank1997fuzzy}
P.~M. Frank and B.~K{\"o}ppen-Seliger, ``Fuzzy logic and neural network
  applications to fault diagnosis,'' \emph{International journal of approximate
  reasoning}, vol.~16, no.~1, pp. 67--88, 1997.

\bibitem{liu2016robust}
J.~Liu, W.~Luo, X.~Yang, and L.~Wu, ``Robust model-based fault diagnosis for
  pem fuel cell air-feed system,'' \emph{IEEE Transactions on Industrial
  Electronics}, vol.~63, no.~5, pp. 3261--3270, 2016.

\bibitem{aldrich2013unsupervised}
C.~Aldrich and L.~Auret, \emph{Unsupervised process monitoring and fault
  diagnosis with machine learning methods}.\hskip 1em plus 0.5em minus
  0.4em\relax Springer, 2013.

\bibitem{lv2016fault}
F.~Lv, C.~Wen, Z.~Bao, and M.~Liu, ``Fault diagnosis based on deep learning,''
  in \emph{2016 American Control Conference (ACC)}.\hskip 1em plus 0.5em minus
  0.4em\relax IEEE, 2016, pp. 6851--6856.

\bibitem{jegadeeshwaran2015fault}
R.~Jegadeeshwaran and V.~Sugumaran, ``Fault diagnosis of automobile hydraulic
  brake system using statistical features and support vector machines,''
  \emph{Mechanical Systems and Signal Processing}, vol.~52, pp. 436--446, 2015.

\bibitem{elforjani2017prognosis}
M.~Elforjani and S.~Shanbr, ``Prognosis of bearing acoustic emission signals
  using supervised machine learning,'' \emph{IEEE Transactions on industrial
  electronics}, vol.~65, no.~7, pp. 5864--5871, 2017.

\bibitem{martin2018experimental}
I.~Martin-Diaz, D.~Morinigo-Sotelo, O.~Duque-Perez, and R.~J. Romero-Troncoso,
  ``An experimental comparative evaluation of machine learning techniques for
  motor fault diagnosis under various operating conditions,'' \emph{IEEE
  Transactions on Industry Applications}, vol.~54, no.~3, pp. 2215--2224, 2018.

\bibitem{verstraete2017deep}
D.~Verstraete, A.~Ferrada, E.~L. Droguett, V.~Meruane, and M.~Modarres, ``Deep
  learning enabled fault diagnosis using time-frequency image analysis of
  rolling element bearings,'' \emph{Shock and Vibration}, vol. 2017, 2017.

\bibitem{glowacz2018early}
A.~Glowacz, W.~Glowacz, Z.~Glowacz, and J.~Kozik, ``Early fault diagnosis of
  bearing and stator faults of the single-phase induction motor using acoustic
  signals,'' \emph{Measurement}, vol. 113, pp. 1--9, 2018.

\bibitem{wang2019novel}
Z.~Wang, J.~Zhou, J.~Wang, W.~Du, J.~Wang, X.~Han, and G.~He, ``A novel fault
  diagnosis method of gearbox based on maximum kurtosis spectral entropy
  deconvolution,'' \emph{IEEE Access}, vol.~7, pp. 29\,520--29\,532, 2019.

\bibitem{zhang2019limited}
A.~Zhang, S.~Li, Y.~Cui, W.~Yang, R.~Dong, and J.~Hu, ``Limited data rolling
  bearing fault diagnosis with few-shot learning,'' \emph{IEEE Access}, vol.~7,
  pp. 110\,895--110\,904, 2019.

\bibitem{gebraeel2009residual}
N.~Gebraeel, A.~Elwany, and J.~Pan, ``Residual life predictions in the absence
  of prior degradation knowledge,'' \emph{IEEE Transactions on Reliability},
  vol.~58, no.~1, pp. 106--117, 2009.

\bibitem{wang2018deep}
M.~Wang and W.~Deng, ``Deep visual domain adaptation: A survey,''
  \emph{Neurocomputing}, vol. 312, pp. 135--153, 2018.

\bibitem{lu2016deep}
W.~Lu, B.~Liang, Y.~Cheng, D.~Meng, J.~Yang, and T.~Zhang, ``Deep model based
  domain adaptation for fault diagnosis,'' \emph{IEEE Transactions on
  Industrial Electronics}, vol.~64, no.~3, pp. 2296--2305, 2016.

\bibitem{guo2018deep}
L.~Guo, Y.~Lei, S.~Xing, T.~Yan, and N.~Li, ``Deep convolutional transfer
  learning network: A new method for intelligent fault diagnosis of machines
  with unlabeled data,'' \emph{IEEE Transactions on Industrial Electronics},
  vol.~66, no.~9, pp. 7316--7325, 2018.

\bibitem{han2019deep}
T.~Han, C.~Liu, W.~Yang, and D.~Jiang, ``Deep transfer network with joint
  distribution adaptation: A new intelligent fault diagnosis framework for
  industry application,'' \emph{ISA transactions}, 2019.

\bibitem{CWRU}
``Case western reserve university bearing data center website,''
  \url{http://csegroups.case.edu/bearingdatacenter/home}.

\bibitem{ponce2006dataset}
J.~Ponce, T.~L. Berg, M.~Everingham, D.~A. Forsyth, M.~Hebert, S.~Lazebnik,
  M.~Marszalek, C.~Schmid, B.~C. Russell, A.~Torralba \emph{et~al.}, ``Dataset
  issues in object recognition,'' in \emph{Toward category-level object
  recognition}.\hskip 1em plus 0.5em minus 0.4em\relax Springer, 2006, pp.
  29--48.

\bibitem{torralba2011unbiased}
A.~Torralba, A.~A. Efros \emph{et~al.}, ``Unbiased look at dataset bias.'' in
  \emph{CVPR}, vol.~1, no.~2.\hskip 1em plus 0.5em minus 0.4em\relax Citeseer,
  2011, p.~7.

\bibitem{tommasi2017deeper}
T.~Tommasi, N.~Patricia, B.~Caputo, and T.~Tuytelaars, ``A deeper look at
  dataset bias,'' in \emph{Domain adaptation in computer vision
  applications}.\hskip 1em plus 0.5em minus 0.4em\relax Springer, 2017, pp.
  37--55.

\bibitem{shimodaira2000improving}
H.~Shimodaira, ``Improving predictive inference under covariate shift by
  weighting the log-likelihood function,'' \emph{Journal of statistical
  planning and inference}, vol.~90, no.~2, pp. 227--244, 2000.

\bibitem{vorburger2006entropy}
P.~Vorburger and A.~Bernstein, ``Entropy-based concept shift detection,'' in
  \emph{Sixth International Conference on Data Mining (ICDM'06)}.\hskip 1em
  plus 0.5em minus 0.4em\relax IEEE, 2006, pp. 1113--1118.

\bibitem{tzeng2014deep}
E.~Tzeng, J.~Hoffman, N.~Zhang, K.~Saenko, and T.~Darrell, ``Deep domain
  confusion: Maximizing for domain invariance,'' \emph{arXiv preprint
  arXiv:1412.3474}, 2014.

\bibitem{gretton2007kernel}
A.~Gretton, K.~Borgwardt, M.~Rasch, B.~Sch{\"o}lkopf, and A.~J. Smola, ``A
  kernel method for the two-sample-problem,'' in \emph{Advances in neural
  information processing systems}, 2007, pp. 513--520.

\bibitem{rozantsev2018beyond}
A.~Rozantsev, M.~Salzmann, and P.~Fua, ``Beyond sharing weights for deep domain
  adaptation,'' \emph{IEEE transactions on pattern analysis and machine
  intelligence}, vol.~41, no.~4, pp. 801--814, 2018.

\bibitem{Tzeng_2017_CVPR}
E.~Tzeng, J.~Hoffman, K.~Saenko, and T.~Darrell, ``Adversarial discriminative
  domain adaptation,'' in \emph{The IEEE Conference on Computer Vision and
  Pattern Recognition (CVPR)}, July 2017.

\bibitem{motiian2017unified}
S.~Motiian, M.~Piccirilli, D.~A. Adjeroh, and G.~Doretto, ``Unified deep
  supervised domain adaptation and generalization,'' in \emph{Proceedings of
  the IEEE International Conference on Computer Vision}, 2017, pp. 5715--5725.

\bibitem{motiian2017few}
S.~Motiian, Q.~Jones, S.~Iranmanesh, and G.~Doretto, ``Few-shot adversarial
  domain adaptation,'' in \emph{Advances in Neural Information Processing
  Systems}, 2017, pp. 6670--6680.

\bibitem{ben2010impossibility}
S.~Ben-David, T.~Lu, T.~Luu, and D.~P{\'a}l, ``Impossibility theorems for
  domain adaptation,'' in \emph{International Conference on Artificial
  Intelligence and Statistics}, 2010, pp. 129--136.

\bibitem{chopra2005learning}
S.~Chopra, R.~Hadsell, Y.~LeCun \emph{et~al.}, ``Learning a similarity metric
  discriminatively, with application to face verification,'' in \emph{CVPR
  (1)}, 2005, pp. 539--546.

\bibitem{bezdek1998multiple}
J.~C. Bezdek, T.~R. Reichherzer, G.~S. Lim, and Y.~Attikiouzel,
  ``Multiple-prototype classifier design,'' \emph{IEEE Transactions on Systems,
  Man, and Cybernetics, Part C (Applications and Reviews)}, vol.~28, no.~1, pp.
  67--79, 1998.

\bibitem{liu2001evaluation}
C.-L. Liu and M.~Nakagawa, ``Evaluation of prototype learning algorithms for
  nearest-neighbor classifier in application to handwritten character
  recognition,'' \emph{Pattern Recognition}, vol.~34, no.~3, pp. 601--615,
  2001.

\bibitem{kohonen1990self}
T.~Kohonen, ``The self-organizing map,'' \emph{Proceedings of the IEEE},
  vol.~78, no.~9, pp. 1464--1480, 1990.

\bibitem{sato1996generalized}
A.~Sato and K.~Yamada, ``Generalized learning vector quantization,'' in
  \emph{Advances in neural information processing systems}, 1996, pp. 423--429.

\bibitem{sato1998formulation}
A.~Sato and K.~Yamada, ``A formulation of learning vector quantization using a new
  misclassification measure,'' in \emph{Proceedings. Fourteenth International
  Conference on Pattern Recognition (Cat. No. 98EX170)}, vol.~1.\hskip 1em plus
  0.5em minus 0.4em\relax IEEE, 1998, pp. 322--325.

\bibitem{snell2017prototypical}
J.~Snell, K.~Swersky, and R.~Zemel, ``Prototypical networks for few-shot
  learning,'' in \emph{Advances in Neural Information Processing Systems},
  2017, pp. 4077--4087.

\bibitem{yang2018robust}
H.-M. Yang, X.-Y. Zhang, F.~Yin, and C.-L. Liu, ``Robust classification with
  convolutional prototype learning,'' in \emph{Proceedings of the IEEE
  Conference on Computer Vision and Pattern Recognition}, 2018, pp. 3474--3482.

\bibitem{zhang2017new}
W.~Zhang, G.~Peng, C.~Li, Y.~Chen, and Z.~Zhang, ``A new deep learning model
  for fault diagnosis with good anti-noise and domain adaptation ability on raw
  vibration signals,'' \emph{Sensors}, vol.~17, no.~2, p. 425, 2017.

\bibitem{hinton2012improving}
G.~E. Hinton, N.~Srivastava, A.~Krizhevsky, I.~Sutskever, and R.~R.
  Salakhutdinov, ``Improving neural networks by preventing co-adaptation of
  feature detectors,'' \emph{arXiv preprint arXiv:1207.0580}, 2012.

\bibitem{zeiler2012adadelta}
M.~D. Zeiler, ``Adadelta: an adaptive learning rate method,'' \emph{arXiv
  preprint arXiv:1212.5701}, 2012.

\bibitem{he2015delving}
K.~He, X.~Zhang, S.~Ren, and J.~Sun, ``Delving deep into rectifiers: Surpassing
  human-level performance on imagenet classification,'' in \emph{Proceedings of
  the IEEE international conference on computer vision}, 2015, pp. 1026--1034.

\bibitem{article}
W.~Smith and R.~Randall, ``Rolling element bearing diagnostics using the case
  western reserve university data: A benchmark study,'' \emph{Mechanical
  Systems and Signal Processing}, vol. 64-65, 05 2015.

\bibitem{zhang2019deep}
M.~Zhang, D.~Wang, W.~Lu, J.~Yang, Z.~Li, and B.~Liang, ``A deep transfer model
  with wasserstein distance guided multi-adversarial networks for bearing fault
  diagnosis under different working conditions,'' \emph{IEEE Access}, vol.~7,
  pp. 65\,303--65\,318, 2019.

\bibitem{maaten2008visualizing}
L.~v.~d. Maaten and G.~Hinton, ``Visualizing data using t-sne,'' \emph{Journal
  of machine learning research}, vol.~9, no. Nov, pp. 2579--2605, 2008.

\end{thebibliography}
%








\end{document}